\documentclass[
reprint,
dcolumn,
bm,
nofootinbib,
notitlepage,
superscriptaddress]{revtex4-2}
\usepackage{lmodern}
\usepackage[latin1]{inputenc}
\usepackage[T1]{fontenc}
\usepackage{graphicx}
\usepackage{xcolor}
\usepackage{amsmath}
\usepackage{amssymb}
\usepackage{braket}
\usepackage{mathrsfs}
\usepackage{mathtools}
\usepackage{hyperref}
\usepackage{ulem}
\usepackage{upgreek}
\usepackage{pstricks-add}
\usepackage{natbib} 
\usepackage{listings}
\usepackage{tikz}
\usepackage{tcolorbox}
\usetikzlibrary{mindmap}
\usepackage{cancel}
\usepackage{algorithm2e}
\usepackage{simpler-wick}
\usepackage{pifont}
\newcommand{\cmark}{\ding{51}}%
\newcommand{\xmark}{\ding{55}}%

\definecolor{lightblue}{rgb}{0.145,0.6666,1}

\begin{document}
\title{Cavity Born-Oppenheimer Coupled Cluster Theory: Towards Electron Correlation in the Vibrational Strong Light-Matter Coupling Regime}

\author{Eric W. Fischer}
\email{ericwfischer.@posteo.de}
\affiliation{Humboldt-Universit\"at zu Berlin, Institut f\"ur Chemie, Brook-Taylor-Stra\ss e 2, D-12489 Berlin, Germany}

\date{\today}

\let\newpage\relax

\begin{abstract}
We present a detailed derivation and discussion of cavity Born-Oppenheimer coupled cluster (CBO-CC) theory and address cavity-modified electron correlation in the vibrational strong coupling regime. Methodologically, we combine the recently proposed cavity reaction potential (CRP) approach with the Lagrangian formulation of CC theory and derive a self-consistent CRP-CC method at the singles and doubles excitations  level (CRP-CCSD). The CRP-CC approach is formally similar to implicit solvation CC models and provides access to the CBO-CC electronic ground state energy minimized in cavity coordinate space on a CC level of theory. A hierarchy of linearisation schemes (lCRP-CCSD) similar to canonical CC theory systematically lifts the self-consistent nature of the CRP-CCSD approach and mitigates numerical cost by approximating electron correlation effects in energy minimization.
We provide a thorough comparison of CRP-CCSD, lCRP-CCSD  and CRP-Hartee-Fock methods for a cavity-modified Menshutkin reaction, pyridine$+$CH$_3$Br, and cavity-induced collective electronic effects in microsolvation energies of selected methanol-water clusters. We find lCRP-CCSD methods to provide excellent results compared to the self-consistent CRP-CCSD approach in the few-molecule limit. We furthermore observe significant differences between mean-field and correlated results in both reactive and collective scenarios. Our work emphasizes the non-trivial character of electron correlation under vibrational strong coupling and provides a starting point for further developments in \textit{ab initio} vibro-polaritonic chemistry beyond the mean-field approximation.
\end{abstract}

\let\newpage\relax
\maketitle
\newpage

\section{Introduction}
\label{sec.intro}
The emerging field of polaritonic chemistry relies on the realization of strong light-matter coupling between spatially confined modes of Fabry-P\'erot cavities and molecular excitations\cite{ebbesen2016,garciavidal2021}, which has been experimentally reported to intriguingly alter both ground and excited state reactivity.\cite{schwartz2011,hutchison2012,shalabney2015,george2015,long2015,thomas2016,thomas2019,ahn2023,patrahau2024} In contrast to traditional light-matter interaction scenarios in chemistry, strong coupling schemes exploit the \textit{quantum} character of light to modify and control chemical reaction pathways. Accordingly, a consistent theoretical approach to strongly coupled light-matter hybrid systems would describe both light and matter degrees of freedom on a quantum mechanical level, which is conceptually realized by the framework of molecular quantum electrodynamics (QED) in the non-relativistic limit.\cite{craig1984,ruggenthaler2023}

In recent years, a significant amount of theoretical research was devoted to the extension of quantum chemical \textit{ab initio} methods to polaritonic chemistry. Conceptually, \textit{ab initio} polaritonic chemistry has to address two dominant flavors of strong light-matter coupling in experimental scenarios: the electronic strong coupling (ESC) and vibrational strong coupling (VSC) regimes. In the ESC regime, high-energy cavity modes couple to electronic excitations, which leads to adiabatic polaritonic states with a mixed fermion-boson character. Theoretically, one employs here a \textit{polaritonic} partitioning scheme, which groups ``fast'' (high-energy) electrons and cavity modes opposed to ``slow'' (low-energy) nuclei.\cite{ruggenthaler2018,rokaj2018} Naturally, the \textit{polaritonic} partitioning scenario is not captured by standard quantum chemistry methods designed for the purely electronic many-body problem. In contrast, the VSC regime is determined by low-energy cavity modes coupled to (ro)vibrational excitations, which leads to the formation of (ro)vibrational polaritons. In this context, the \textit{cavity Born-Oppenheimer} (CBO) partitioning has been proposed, which extends the molecular Born-Oppenheimer scheme by combining ``slow'' cavity modes and nuclei while electrons are treated as ``fast'' degrees of freedom.\cite{flick2017,flick2017cbo,fischer2023} The CBO scheme leads to purely electronic CBO adiabatic states and energies, which both however exhibit an additional parametric dependence on cavity displacement coordinates manifesting in the concept of cavity potential energy surfaces (cPES). 

Methodologically, there exist two paradigms in \textit{ab initio} polaritonic chemistry in analogy to quantum chemistry: Density-based methods as realized by quantum electrodynamical density functional theory (QEDFT)\cite{ruggenthaler2011,tokatly2013,ruggenthaler2014,flick2015,ruggenthaler2017} and wave function based approaches, which will be the focus of this work. \textit{Ab initio} wave function approaches to the ESC regime were realized via extensions of established (electronic) quantum chemical methods to the mixed fermion-boson problem: QED Hartree-Fock theory (QED-HF)\cite{haugland2020}, its strong-coupling (SC-QED-HF) and Lang-Firsov variants (LF-HF)\cite{riso2022,cui2024}, configuration interaction theory (QED-CI)\cite{haugland2021,mctague2022}, coupled cluster theory (QED-CC) and its equation-of-motion extension (QED-EOM-CC)\cite{haugland2020,deprince2021,monzel2024}, M\o ller-Plesst perturbation theory\cite{cui2024,bauer2023,elmoutaoukal2025} as well as more recent adaptations of multireference methods like the  complete active space configuration interaction (QED-CASCI)\cite{vu2024}, density matrix renormalization group (QED-DMRG)\cite{matousek2024} and complete active space self-consistent field (QED-CASSCF)\cite{alessandro2025,vu2025} approaches. 

In contrast, the VSC regime is significantly less studied from an \textit{ab initio} perspective. Only recently Schnappinger \textit{et al.} introduced the CBO Hartree-Fock (CBO-HF) method in a restricted formulation, which constitutes a mean-field approach to electronic interactions of closed-shell systems under VSC.\cite{schnappinger2023} Those authors additionally connected minimization of the CBO electronic ground state energy in cavity coordinate space to the nonradiating ground state condition for the transverse electric cavity field.\cite{schnappinger2023} We showed in a subsequent study that the nonradiating ground state condition can be directly accounted for in a nonlinear reformulation of the \textit{exact} CBO electronic ground state problem.\cite{fischer2024} This cavity reaction potential (CRP) approach directly provides access to the CBO electronic ground state energy minimized in cavity coordinate space without the need of explicit cavity gradients. We subsequently proposed extensions of the CBO-HF method and the CBO coupled cluster (CBO-CC) approach, the latter previously discussed only for the ESC regime\cite{angelico2023}, to the CRP framework.\cite{fischer2024} First results obtained via the CRP-CC approach theoretically illustrated the non-trivial role of electron correlation effects under VSC not captured by mean-field theory.\cite{fischer2024}

In this contribution, we will now provide a detailed derivation and discussion of both CBO-HF and CBO-CC approaches in the CRP framework, which were missing so far. The CRP-CC method is considered at the singles and doubles excitation level (CRP-CCSD), and relies on a Lagrangian formulation formally similar to implicit solvation CC models\cite{christiansen1999,cammi2009,caricato2010,caricato2011}. In this context, a self-consistent solution of the correlated CBO electronic problem is necessary, which encodes electronic energy optimization in cavity coordinate space at a CC level of theory. We furthermore introduce a hierarchy of approximations inspired by implicit solvation CC theory\cite{cammi2009,caricato2011}, which lift the self-consistent nature and lead to linearized CRP-CC (lCRP-CC) methods similar to canonical CC theory. A comparison of different (approximate) \textit{ab initio} CRP-CCSD methods illustrate their capabilities and the relevance of electron correlation under VSC for selected examples of cavity-modified reaction energies and collective effects in microsolvation energies under VSC.

The paper is structured as follows. In Sec.\ref{sec.theory_background}, we discuss the CBO electronic ground state problem and its CRP formulation followed by a derivation of the CRP-HF method in Sec.\ref{sec.crp_rhf}. In Sec.\ref{sec.crp_cc_theory}, we introduce the self-consistent CRP-CCSD approach and its implementation followed by approximate lCRP-CCSD schemes in Sec.\ref{sec.linear_schemes}. We compare CRP-CCSD, lCRP-CCSD and CRP-HF methods illustratively for selected molecular examples in Sec.\ref{sec.num_exp}. Sec.\ref{sec.conclusion} concludes this work.

\section{Theoretical Background}
\label{sec.theory_background}
We introduce the CBO electronic ground state problem and its CRP formulation, which provides the conceptual basis for the remainder of this work.

\subsection{The CBO Electronic Ground State Problem}
We consider a molecular subsystem strongly coupled to a single effective mode of an infrared optical cavity in the dipole approximation. In length gauge representation, the adiabatic ground state cavity potential energy surface (cPES) is given by\cite{flick2017cbo,fischer2023}
\begin{align}
E^{(ec)}_0(\underline{R},x_\lambda)
&=
\braket{
\Psi^{(ec)}_0(\underline{R},x_\lambda)
\vert
\hat{H}_{ec}
\vert
\Psi^{(ec)}_0(\underline{R},x_\lambda)}
\quad,
\label{eq.ground_state_cpes}
\end{align}
with CBO adiabatic ground state, $\ket{\Psi^{(ec)}_0(\underline{R},x_\lambda)}$, which parametrically depends on both nuclear, $\underline{R}$, and cavity displacement coordinates, $x_\lambda$, for a cavity mode with polarization, $\lambda$. We assume a valid CBO approximation, \textit{i.e.}, $E^{(ec)}_0$ is energetically well separated from the excited state manifold, such that non-adiabatic effects are negligibly small. 

The CBO electronic Hamiltonian, $\hat{H}_{ec}$, in Eq.\eqref{eq.ground_state_cpes} is given in second quantization representation as\cite{angelico2023,fischer2024}
\begin{align}
\hat{H}_{ec}
&=
\sum_{pq}
h^{pq}_\lambda
\hat{E}_{pq}
+
\dfrac{1}{2}
\sum_{pqrs}
\tilde{g}_{pqrs}
\hat{e}_{pqrs}
+
\tilde{V}_{nc}
\quad,
\label{eq.cbo_electronic_hamilton}
\end{align}
where we employ the common notation for general ($p,q,r,s$), occupied ($i,j,k,l$) and virtual ($a,b,c,d$) molecular orbital (MO) indices. Spin-summed electronic one- and two-particle excitation operators are explicitly given by
\begin{align}
\hat{E}_{pq}
&=
\sum_\sigma
\hat{a}^\dagger_{p\sigma}
\hat{a}_{q\sigma}
\quad,
\vspace{0.2cm}
\\
\hat{e}_{pqrs}
&=
\hat{E}_{pq}
\hat{E}_{rs}
-
\delta_{qr}
\hat{E}_{ps}
\quad,
\end{align}
with spin index $\sigma$. The CBO one- and two-electron integrals, $h^{pq}_\lambda$ and $\tilde{g}_{pqrs}$, take the form
\begin{align}
h^{pq}_\lambda
&=
h_{pq}
+
\dfrac{g^2_0}{2}
O^{pq}_\lambda
-
g_0
\omega_c
d^{pq}_\lambda
x_\lambda
+
g^2_0
d^{pq}_\lambda
\hat{d}^{(n)}_\lambda
\quad,
\label{eq.cbo_core}
\vspace{0.2cm}
\\
\tilde{g}_{pqrs}
&=
(pq\vert rs)
+
g^2_0
d^{pq}_\lambda
d^{rs}_\lambda
\quad,
\label{eq.cbo_eri}
\end{align} 
with electron repulsion integrals, $(pq\vert rs)$, in Mulliken notation and polarization-projected electronic dipole matrix elements
\begin{align}
d^{pq}_\lambda
=
-
e
\braket{
\chi_p
\vert
r_{i\lambda}
\vert
\chi_q
}
\quad.
\end{align}
The one-electron term, $h^{pq}_\lambda$, contains the canonical core contribution, $h_{pq}$, a one-electron DSE term determined by a matrix element of the polarization-projected electronic quadrupole moment
\begin{align}
O^{pq}_\lambda
=
e^2
\braket{
\chi_p
\vert
r^2_{i\lambda}
\vert
\chi_q
}
\quad,
\end{align}
the cavity-electron interaction term with harmonic cavity frequency, $\omega_c$, and a DSE cross term, which couples electrons and nuclei via the polarization projected nuclear dipole operator, $\hat{d}^{(n)}_\lambda$, respectively. The light-matter coupling constant, $g_0=\frac{1}{\sqrt{\epsilon_0\,V_c}}$, is proportional to the cavity mode volume, $V_c$, but will be treated as a free parameter in the following. Finally, the remaining term of the CBO electronic Hamiltonian is a nuclear-cavity contribution 
\begin{align}
\tilde{V}_{nc}
&=
V_{nn}
+
\dfrac{\omega^2_c}{2}
x^2_\lambda
-
g_0
\omega_c
\hat{d}^{(n)}_\lambda
x_\lambda
+
\dfrac{g^2_0}{2}
\hat{d}^{(n)}_\lambda
\hat{d}^{(n)}_\lambda
\quad,
\end{align} 
which contains nuclear interaction and harmonic cavity potentials, the nuclei-cavity interaction besides a purely nuclear DSE term, and constitutes therefore a constant energy shift of the CBO electronic energy.

\subsection{The Ground State Cavity Reaction Potential}
We will now expand the ground state cPES in Eq.\eqref{eq.ground_state_cpes} around a stationary configuration, $(\underline{R},x^0_\lambda)$, in a second-order Taylor series\cite{fischer2024}
\begin{align}
E^{(ec)}_0(\underline{R},x_\lambda)
&=
\mathcal{V}^{(ec)}_0(\underline{R})
+
\dfrac{1}{2}
\underline{C}^T
\underline{\underline{H}}^{(ec)}_0
\underline{C}
\quad,
\label{eq.taylor_cpes}
\end{align}
and denote the first term as ground state cavity reaction potential (CRP)
\begin{align}
\mathcal{V}^{(ec)}_0(\underline{R})
=
E^{(ec)}_0(\underline{R},x^0_\lambda)
\quad,
\label{eq.crp_def}
\end{align}
which is simply the ground state cPES evaluated at the stationary cavity coordinate, $x^0_\lambda$. The second term in Eq.\eqref{eq.taylor_cpes} contains the vibro-polaritonic Hessian, $\underline{\underline{H}}^{(ec)}_0$, which has been discussed previously from both non-perturbative and perturbative perspectives.\cite{bonini2022,schnappinger2023ir,fischer2024ir} Since the CBO electronic Hamiltonian in Eq.\eqref{eq.cbo_electronic_hamilton} is at most quadratic in cavity displacement coordinates, the expansion in Eq.\eqref{eq.taylor_cpes} is formally exact for the cavity subsystem as all higher-order terms in $x_\lambda$ vanish identically. 

The expression in Eq.\eqref{eq.crp_def} generalizes an earlier idea from effective vibro-polaritonic reaction models\cite{fischer2022} to \textit{ab initio} vibro-polaritonic chemistry: $\mathcal{V}^{(ec)}_0$ can be interpreted as a minimum energy surface in cavity coordinate space that fully captures static properties of cavity-modified chemical reactions. Inspired by concepts from reaction rate theory\cite{miller1980}, this motivates the notion of a cavity reaction potential. The second term in Eq.\eqref{eq.taylor_cpes} provides a harmonic potential perpendicular to $\mathcal{V}^{(ec)}_0$ that accounts for excitations of the strongly coupled nuclear-cavity subsystem, where we treat the nuclei here in harmonic approximation.

We shall now inspect the CRP as defined in Eq.\eqref{eq.crp_def} in more detail. The stationary cavity coordinate, $x^0_\lambda$, minimizes the CBO electronic energy in cavity coordinate space and equivalently satisfies the nonradiating ground state condition\cite{schnappinger2023,fischer2024}
\begin{align}
\left.
\dfrac{\partial}{\partial x_\lambda}
E^{(ec)}_0
\right\vert_{x^0_\lambda}
=
0
\,
\Leftrightarrow
\,
\braket{
\Psi^{(ec)}_0
\vert
\underline{\hat{E}}_\perp
\vert
\Psi^{(ec)}_0}
=
0
\quad,
\label{eq.crp_constraints}
\end{align}
where $\underline{\hat{E}}_\perp$ is the transverse electric field operator of the cavity field. One can exploit the Hellmann-Feynman theorem to evaluate the stationary cavity coordinate exactly as
\begin{align}
x^0_\lambda 
&=
\dfrac{g_0}{\omega_c}
\braket{
\Psi^{(ec)}_0
\vert 
\hat{d}^{(en)}_\lambda 
\vert
\Psi^{(ec)}_0}
\quad,
\label{eq.mincav_exact}
\end{align}
with polarization-projected molecular dipole operator, $\hat{d}^{(en)}_\lambda=\hat{d}^{(e)}_\lambda+\hat{d}^{(n)}_\lambda$, composed of electronic ($\hat{d}^{(e)}_\lambda$) and nuclear ($\hat{d}^{(n)}_\lambda$) components, respectively. We note here that due to the quadratic character of $\hat{H}_{ec}$ in terms of $x_\lambda$, $x^0_\lambda$ always refers to a minimum. From Eqs.\eqref{eq.ground_state_cpes} and \eqref{eq.mincav_exact}, we obtain an explicit expression for the ground state CRP (\textit{cf.} Appendix \ref{subsec.derivation_crp})
\begin{multline}
\mathcal{V}^{(ec)}_0(\underline{R})
=
\braket{
\Psi^{(ec)}_0
\vert
\hat{\mathcal{H}}_e
\vert
\Psi^{(ec)}_0}
\\
-
\dfrac{g^2_0}{2}
\braket{
\Psi^{(ec)}_0
\vert
\hat{d}^{(e)}_\lambda
\vert
\Psi^{(ec)}_0}^2
\quad,
\label{eq.exact_crp}
\end{multline} 
which is independent of both the nuclear dipole contribution and the cavity frequency, origin invariant (\textit{cf.} Appendix \ref{subsec.origin_invariance_crp}) and subject to an effective CBO electronic Hamiltonian 
\begin{align}
\hat{\mathcal{H}}_e
&=
\sum_{pq}
\tilde{h}^{pq}_\lambda
\hat{E}_{pq}
+
\dfrac{1}{2}
\sum_{pqrs}
\tilde{g}_{pqrs}
\hat{e}_{pqrs}
+
V_{nn}
\quad,
\label{eq.eff_hamilton}
\end{align}
with DSE-augmented core contribution
\begin{align}
\tilde{h}^{pq}_\lambda
&=
h_{pq}
+
\dfrac{g^2_0}{2}
O^{pq}_\lambda
\quad,
\label{eq.hcore_dse}
\end{align} 
and CBO electron repulsion integral, $\tilde{g}_{pqrs}$ as given in Eq.\eqref{eq.cbo_eri}.

We shall now exploit the observation that Eq.\eqref{eq.exact_crp} can be interpreted as a ground state-projected nonlinear electronic Schr\"odinger equation 
\begin{align}
\mathcal{V}^{(ec)}_0
&=
\braket{
\Psi^{(ec)}_0
\vert
\hat{\mathcal{H}}^{\Psi_0}_e
\vert
\Psi^{(ec)}_0}
\quad,
\label{eq.exact_crp_nonlinear}
\end{align} 
specified by a respectively nonlinear CBO electronic Hamiltonian
\begin{align}
\hat{\mathcal{H}}^{\Psi_0}_e
&=
\hat{\mathcal{H}}_e
-
\dfrac{g^2_0}{2}
\braket{
\Psi^{(ec)}_0
\vert
\hat{d}^{(e)}_\lambda
\vert
\Psi^{(ec)}_0}
\hat{d}^{(e)}_\lambda
\quad.
\label{eq.exact_nonlinear_eff_hamilton}
\end{align}
The nonlinearity emerges from the second term, which explicitly depends on the one-particle reduced density matrix (1-RDM) determining the electronic dipole expectation value. Moreover, Eq.\eqref{eq.exact_crp_nonlinear} provides access to the ground state CBO electronic energy minimized in cavity coordinate space without the need of cavity coordinate gradients. However, this gain comes at the cost of Eq.\eqref{eq.exact_crp_nonlinear} being nonlinear, which requires a self-consistent solution process that resembles energy minimization in cavity coordinate space.

At this point, we shall discuss why one should aim at solving the nonlinear CRP ground state problem in Eq.\eqref{eq.exact_crp_nonlinear} instead of Eq.\eqref{eq.ground_state_cpes} followed by a common quantum chemical optimization routine extended to cavity modes. First, both ground state energy and wave function correspond to a chemically meaningful stationary point on the cPES, which can be analyzed with respect to reaction mechanism and cavity-modified electron correlation. Second, one circumvents the need of extending analytic gradients at different levels of theory to cavity coordinate space. Third, Hartree-Fock theory as reference for correlated wave function methods in quantum chemistry is straightforwardly generalizable to the CRP scenario due to its inherently nonlinear character at no additional cost. Forth, it turns out that an approximation to the CRP problem in terms of coupled cluster theory is formally similar to implicit solvation coupled cluster theory, which allows us here to benefit from an established theoretical framework transferred to a different context.
In the remainder of this work, we will discuss approximations to Eq.\eqref{eq.exact_crp_nonlinear} based on Hartree-Fock and coupled cluster theory.

\section{CRP Hartree-Fock Theory}
\label{sec.crp_rhf}
We introduce the CRP formulation of CBO Hartree-Fock (CRP-HF) theory, which resembles a mean-field approximation of the ground state CRP. It furthermore serves as a starting point for the development of a related CRP coupled cluster (CPR-CC) method to be discussed in Sec.\ref{sec.crp_cc_theory}. 

The CRP-HF approach relies on a single-determinant approximation of the CBO adiabatic ground state 
\begin{align}
\ket{\Psi^{(ec)}_0}
\approx
e^{-\hat{\kappa}}
\ket{\Phi^{(ec)}_0}
=
\ket{\Phi^{(ec)}_0(\underline{\kappa})}
\quad,
\end{align} 
which turns Eq.\eqref{eq.exact_crp} into
\begin{multline}
\mathcal{V}^{(ec)}_\mathrm{rhf}(\underline{\kappa})
=
\braket{
\Phi^{(ec)}_0(\underline{\kappa})
\vert
\hat{\mathcal{H}}_e
\vert
\Phi^{(ec)}_0(\underline{\kappa})}
\\
-
\dfrac{g^2_0}{2}
\braket{
\Phi^{(ec)}_0(\underline{\kappa})
\vert
\hat{d}^{(e)}_\lambda
\vert
\Phi^{(ec)}_0(\underline{\kappa})}^2
\quad.
\label{eq.crp_rhf_energy}
\end{multline} 
Here, we introduced the one-electron orbital rotation operator\cite{helgakerbook}
\begin{align}
\hat{\kappa}
=
\sum_{p>q}
\kappa_{pq}
\left(
\hat{E}_{pq}
-
\hat{E}_{qp}
\right)
=
\sum_{p>q}
\kappa_{pq}
\hat{E}^-_{pq}
\quad,
\end{align}
which is parametrically dependent on orbital rotation parameters, $\underline{\kappa}=(\dots\kappa_{pq}\dots)$. In order to minimize the energy in Eq.\eqref{eq.crp_rhf_energy} with respect to orbital rotations, we require the CRP orbital gradient to vanish (\textit{cf.} Appendix \ref{subsec.derivation_crprhf})
\begin{align}
\mathcal{V}^{(1)}_{pq}
=
\braket{
\Phi^{(ec)}_0
\vert
[
\hat{E}^-_{pq},
\hat{\mathcal{H}}^{\Phi_0}_e
]
\vert
\Phi^{(ec)}_0}
=
0
\quad,
\label{eq.crp_orbital_gradient}
\end{align}
where $\hat{\mathcal{H}}^{\Phi_0}_e$ is the mean-field approximation of Eq.\eqref{eq.exact_nonlinear_eff_hamilton} given by 
\begin{align}
\hat{\mathcal{H}}^{\Phi_0}_e
&=
\hat{\mathcal{H}}_e
-
\dfrac{g^2_0}{2}
\braket{
\Phi^{(ec)}_0
\vert
\hat{d}^{(e)}_\lambda
\vert
\Phi^{(ec)}_0}
\hat{d}^{(e)}_\lambda
\quad.
\label{eq.crprhf_nonlinear_eff_hamilton}
\end{align}
The CRP-Fock operator follows as
\begin{align}
\hat{f}^{(e)}_\lambda
&=
\sum_{pq}
\left(
\tilde{h}^{pq}_\lambda
+
\tilde{v}^{pq}_\lambda
\right)
\hat{E}_{pq}
\quad,
\label{eq.crp_fock}
\end{align}
where, $\tilde{h}^{pq}_\lambda$, is given by Eq.\eqref{eq.hcore_dse} and the effective potential reads in the restricted scenario
\begin{align}
\tilde{v}^{pq}_\lambda
&=
\sum_i
\left(
2g_{pqii}
-
g_{piiq}
-
g^2_0
d^{pi}_\lambda
d^{iq}_\lambda
\right)
\quad,
\label{eq.crp_fock_pot}
\end{align}
with canonical Coulomb ($g_{pqii}$) and exchange integrals ($g_{piiq}$), respectively. We note that Eq.\eqref{eq.crp_fock_pot} contains only a DSE-induced exchange contribution while a Coulomb-like equivalent vanishes in the CRP formulation of CBO-HF theory.\cite{fischer2024}

In the canonical MO basis, which diagonalizes the CRP-Fock operator, the mean-field CRP turns into (\textit{cf.} Appendix \ref{subsec.derivation_crprhf})
\begin{align}
\mathcal{V}^{(ec)}_\mathrm{rhf}
=
E^{(ec)}_\mathrm{rhf}
+
g^2_0
\left(
\sum_i
O^{ii}_\lambda
-
\sum_{ij}
d^{ij}_\lambda
d^{ji}_\lambda
\right)
\quad,
\label{eq.mean_field_crp}
\end{align}
where the first term is formally identical to the RHF energy but here evaluated with respect to DSE-relaxed MOs. The remaining two terms reflect a quadrupole- and exchange-type correction emerging from the electronic dipole self-energy and render $\mathcal{V}^{(ec)}_\mathrm{rhf}$ origin invariant. We note that the CRP-HF approach and the CBO-HF method augmented by a minimization routine in cavity coordinate space give identical energies, however, the CRP-HF approach requires only a single self-consistent field cycle to arrive at the stationary result.

\section{CRP Coupled Cluster Theory}
\label{sec.crp_cc_theory}
We are now in the position to discuss the CRP approach in the context of coupled cluster theory, which will allow us to address electron correlation in the VSC regime. In the following derivation, we exploit conceptual ideas from implicit solvation CC theory.\cite{christiansen1999,cammi2009}

\subsection{The CRP Coupled Cluster Lagrangian}
We approximate the CBO adiabatic ground state in Eq.\eqref{eq.exact_crp} by CC states
\begin{align}
\ket{\Psi^{(ec)}_0}
&\approx
\ket{\Psi^{(ec)}_\mathrm{CC}}
=
e^{\hat{T}}
\ket{\Phi^{(ec)}_0}
\quad,
\vspace{0.2cm}
\\
\bra{\Psi^{(ec)}_0}
&\approx
\bra{\Psi^{(ec)}_\Lambda}
=
\bra{\Phi^{(ec)}_0}
(
1
+
\hat{\Lambda}
)
e^{-\hat{T}}
\quad,
\end{align}
with CRP-HF determinant, $\ket{\Phi^{(ec)}_0}$, and introduce the CRP-CC Lagrangian 
\begin{align}
\mathcal{L}^{(ec)}_\mathrm{cc}
=
\braket{
\Psi^{(ec)}_\Lambda
\vert
\hat{\mathcal{H}}_e
\vert
\Psi^{(ec)}_\mathrm{CC}}
-
\dfrac{g^2_0}{2}
\braket{
\Psi^{(ec)}_\Lambda
\vert
\hat{d}^{(e)}_\lambda
\vert
\Psi^{(ec)}_\mathrm{CC}}^2
\,.
\label{eq.crp_cbo_cc_lagrangian}
\end{align}
The CRP-CC Lagrangian depends on cluster and de-excitation operators
\begin{align}
\hat{T}
=
\hat{T}_1
+
\hat{T}_2
\quad,
\quad
\hat{\Lambda}
=
\hat{\Lambda}_1
+
\hat{\Lambda}_2
\quad,
\end{align}
which we restrict in this work to the singles and doubles excitation level with 
\begin{align}
\hat{T}_1
&=
\sum_{ai}
t^a_i
\hat{E}_{ai}
\quad,
\,
\hat{T}_2
=
\dfrac{1}{4}
\sum_{aibj}
t^{ab}_{ij}
\hat{E}_{ai}
\hat{E}_{bj}
\quad,
\vspace{0.2cm}
\\
\hat{\Lambda}_1
&=
\sum_{ai}
\lambda^i_a
\hat{E}^\dagger_{ai}
\quad,
\,
\hat{\Lambda}_2
=
\dfrac{1}{4}
\sum_{aibj}
\lambda^{ij}_{ab}
\hat{E}^\dagger_{ai}
\hat{E}^\dagger_{bj}
\quad,
\end{align}
where we have singles and doubles cluster amplitudes, $t^{a}_{i},t^{ab}_{ij}$, and corresponding multipliers, $\lambda^{i}_{a},\lambda^{ij}_{ab}$, respectively. We consider now the definition of a normal-ordered electronic operator with respect to the CRP-HF reference state
\begin{align}
\{\hat{O}\}
=
\hat{O}
-
\braket{
\Phi^{(ec)}_0
\vert
\hat{O}
\vert
\Phi^{(ec)}_0}
\quad,
\end{align}
which allows us to introduce the normal-ordered CRP-CC Lagrangian as (\textit{cf.} Appendix \ref{subsec.derivation_normal_crp_cc_lagrange})
\begin{multline}
\mathcal{L}^{(ec)}_\mathrm{cc}
=
\mathcal{V}^{(ec)}_\mathrm{rhf}
+
\braket{
\Psi^{(ec)}_\Lambda
\vert
\{\hat{\mathcal{H}}^\mathrm{crp}_e\}
\vert
\Psi^{(ec)}_\mathrm{CC}}
\\
-
\dfrac{g^2_0}{2}
\braket{
\Psi^{(ec)}_\Lambda
\vert
\{\hat{d}^{(e)}_\lambda\}
\vert
\Psi^{(ec)}_\mathrm{CC}}^2
\quad,
\label{eq.normal_ordered_crp_cc_lagrangian}
\end{multline}
where the first term resembles the mean-field CRP in Eq.\eqref{eq.mean_field_crp}. In the second term, we introduced a normal-ordered effective Hamiltonian
\begin{align}
\{\mathcal{\hat{H}}^\mathrm{crp}_e\}
=
\{\hat{f}^{(e)}_\lambda\}
+
\{\hat{W}_{ee}\}
\quad,
\label{eq.crp_eff_hamilton}
\end{align}
with normal-ordered CRP-Fock operator, $\{\hat{f}^{(e)}_\lambda\}$ (\textit{cf.} Eq.\eqref{eq.crp_fock}), and CBO two-electron interaction
\begin{align}
\{\hat{W}_{ee}\}
&=
\dfrac{1}{4}
\sum_{pqrs}
\bar{w}^{pq}_{rs}
\{\hat{e}_{pqrs}\}
\quad,
\end{align}
where the anti-symmetrized CBO electron repulsion integral is given by $\bar{w}^{pq}_{rs}=\tilde{g}_{pqrs}-\tilde{g}_{psrq}$ (\textit{cf.} Eq.\eqref{eq.cbo_eri}). The last term in Eq.\eqref{eq.normal_ordered_crp_cc_lagrangian} contains the correlated contribution of the CC electronic dipole expectation value and is nonlinear in $\Lambda$-multipliers. This nonlinearity is absent in canonical CC theory, where the Lagrangian is linear in $\Lambda$-multipliers, and leads here to coupled working equations in analogy to implicit solvation CC models\cite{christiansen1999,cammi2009}.

\subsection{CRP-CC Equations}
\label{subsec.crp_cc_equaitons}
We obtain a consistent representation of the CRP-CC equations from a slightly rewritten normal-ordered CRP-CC Lagrangian 
\begin{multline}
\mathcal{L}^{(ec)}_\mathrm{cc}
=
\mathcal{V}^{(ec)}_\mathrm{rhf}
+
\braket{
\Psi^{(ec)}_\Lambda
\vert
\{\hat{\mathcal{H}}^\Lambda_e\}
\vert
\Psi^{(ec)}_\mathrm{CC}}
\\
+
\dfrac{g^2_0}{2}
\braket{
\Psi^{(ec)}_\Lambda
\vert
\{\hat{d}^{(e)}_\lambda\}
\vert
\Psi^{(ec)}_\mathrm{CC}}^2
\quad,
\label{eq.lambda_normal_ordered_crp_cc_lagrangian}
\end{multline} 
where we introduced a nonlinear effective Hamiltonian
\begin{align}
\{\hat{\mathcal{H}}^\Lambda_e\}
=
\{\hat{\mathcal{H}}^\mathrm{crp}_e\}
-
g^2_0
\braket{
\Psi^{(ec)}_\Lambda
\vert
\{\hat{d}^{(e)}_\lambda\}
\vert
\Psi^{(ec)}_\mathrm{CC}}
\{\hat{d}^{(e)}_\lambda\}
\quad,
\label{eq.lambda_crp_eff_hamilton}
\end{align}
which can be interpreted as normal-ordered CC approximation of Eq.\eqref{eq.exact_nonlinear_eff_hamilton}. Variation of Eq.\eqref{eq.lambda_normal_ordered_crp_cc_lagrangian} with respect to both multipliers and amplitudes
\begin{align}
\dfrac{\partial \mathcal{L}^{(ec)}_\mathrm{cc}}{\partial \lambda_\nu}
=
0
\quad,
\quad
\dfrac{\partial \mathcal{L}^{(ec)}_\mathrm{cc}}{\partial t_\nu}
=
0
\quad,
\end{align}
leads to compact expressions for the CRP-CC equations
\begin{align}
\braket{
\Phi_\nu
\vert
e^{-\hat{T}}
\{\hat{\mathcal{H}}^\Lambda_e\}
e^{\hat{T}}
\vert
\Phi_0}
=
0
\quad,
\label{eq.amplitudes_crp_cc}
\vspace{0.2cm}
\\
\braket{
\Phi_0
\vert
(
1
+
\hat{\Lambda}
)
[
e^{-\hat{T}}
\{\hat{\mathcal{H}}^\Lambda_e\}
e^{\hat{T}}
,
\hat{\tau}_\nu
]
\vert
\Phi_0}
=
0
\quad,
\label{eq.multipliers_crp_cc}
\end{align} 
with excited determinants, $\ket{\Phi_\nu}=\hat{\tau}_\nu\ket{\Phi_0}$, generated via an excitation operator, $\hat{\tau}_\nu$, acting on the CRP-RHF reference state, $\ket{\Phi_0}$. We realize here that both CPR-CC amplitude and multiplier equations depend on $\Lambda$-multipliers due to the definition of $\{\mathcal{\hat{H}}^\Lambda_e\}$ in Eq.\eqref{eq.lambda_crp_eff_hamilton}, which requires a self-consistent solution procedure. The nonlinear effective Hamiltonian is explicitly given by
\begin{align}
\{\mathcal{\hat{H}}^\Lambda_e\}
=
\{\hat{f}^\Lambda_\lambda\}
+
\{\hat{W}_{ee}\}
\quad,
\label{eq.lambda_eff_hamilton}
\end{align}
with nonlinear CRP-Fockian 
\begin{align}
\{\hat{f}^\lambda_\lambda\}
&=
\{\hat{f}^{(e)}_\lambda\}
-
g^2_0
\sum_{rs}
d^{rs}_\lambda
\gamma^\Lambda_{rs}
\sum_{pq}
d^{pq}_\lambda
\{\hat{E}_{pq}\}
\quad,
\label{eq.lambda_eff_fock}
\end{align}
where the second term depends on the CC response 1-RDM\cite{cammi2009}
\begin{align}
\gamma^\Lambda_{rs}
&=
\braket{
\Phi_0
\vert
(
1
+
\hat{\Lambda}
)
e^{-\hat{T}}
\{\hat{E}_{rs}\}
e^{\hat{T}}
\vert
\Phi_0}
\quad,
\end{align}
with elements\cite{cammi2009}
\begin{align}
\begin{matrix*}[l]
\gamma^\Lambda_{ij}
=
-
\displaystyle\sum_c
\lambda^j_c
t^c_i
-
\dfrac{1}{2}
\displaystyle\sum_{kcd}
\lambda^{jk}_{cd}
t^{cd}_{ik}
\vspace{0.2cm}
\\
\gamma^\Lambda_{ab}
=
\displaystyle\sum_k
\lambda^k_b
t^a_k
+
\dfrac{1}{2}
\displaystyle\sum_{klc}
\lambda^{kl}_{bc}
t^{ac}_{kl}
\vspace{0.2cm}
\\
\gamma^\Lambda_{ia}
=
\lambda^i_a
\vspace{0.2cm}
\\
\gamma^\Lambda_{ai}
=
t^a_i
+
\displaystyle\sum_{jb}
\lambda^j_b
(
t^{ba}_{ji}
-
t^b_i
t^a_j
)
-
\dfrac{1}{2}
\displaystyle\sum_{jkcb}
\lambda^{kj}_{cb}
(
t^{cb}_{ki}
t^a_j
-
t^{ca}_{kj}
t^b_i
)
\end{matrix*}
\,.
\label{eq.response_1rdm}
\end{align}
In Sec.\ref{sec.linear_schemes}, we discuss how a linearisation of the CRP-CC Lagrangian with respect to $\Lambda$-multipliers decouples the CRP-CC equations and allows for systematically approximating energy optimization in cavity coordinate space. 

\subsection{CRP-CC Correlation Energy}
An explicit expression for the CRP-CC correlation energy is obtained from the normal-ordered CRP-CC Lagrangian in Eq.\eqref{eq.lambda_normal_ordered_crp_cc_lagrangian}, which can be equivalently written as (\textit{cf.} Appendix \ref{subsec.alterantive_normal_crp_cc_lagrange})
\begin{align}
\mathcal{L}^{(ec)}_\mathrm{cc}
=
\mathcal{V}^{(ec)}_\mathrm{cc}
+
\sum_\nu
\lambda_\nu
\braket{
\Phi_\nu
\vert
e^{-\hat{T}}
\{\hat{\mathcal{H}}^\Lambda_e\}
e^{\hat{T}}
\vert
\Phi_0}
\quad.
\label{eq.lagrangian_crp_energy}
\end{align}
The second term resembles the CRP-CC amplitude equations \eqref{eq.amplitudes_crp_cc}, which vanish for converged calculations and the first term corresponds to the CRP-CC energy
\begin{align}
\mathcal{V}^{(ec)}_\mathrm{cc}
=
\mathcal{V}^{(ec)}_\mathrm{rhf}
+
\Delta
\mathcal{V}^{(ec)}_\mathrm{cc}
\quad,
\end{align}
which decomposes into the CRP-RHF energy in Eq.\eqref{eq.mean_field_crp} and a CRP-CC correlation correction
\begin{align}
\Delta\mathcal{V}^{(ec)}_\mathrm{cc}
&=
\Delta\mathcal{V}^{\mathcal{H}}_\mathrm{cc}
+
\Delta\mathcal{V}^{d}_\mathrm{cc}
+
\Delta\mathcal{V}^{\Lambda}_\mathrm{cc}
\quad,
\label{eq.exact_crp_cc_energy}
\end{align}
with
\begin{align}
\Delta\mathcal{V}^{\mathcal{H}}_\mathrm{cc}
&=
\braket{
\Phi_0
\vert
e^{-\hat{T}}
\{\hat{\mathcal{H}}^\mathrm{crp}_e\}
e^{\hat{T}}
\vert
\Phi_0}
\quad,
\label{eq.l0_crp_cc_energy}
\vspace{0.2cm}
\\
\Delta\mathcal{V}^{d}_\mathrm{cc}
&=
-
\dfrac{g^2_0}{2}
\braket{
\Phi_0
\vert
e^{-\hat{T}}
\{\hat{d}^{(e)}_\lambda\}
e^{\hat{T}}
\vert
\Phi_0}^2
\quad,
\label{eq.l1_crp_cc_energy}
\vspace{0.2cm}
\\
\Delta\mathcal{V}^{\Lambda}_\mathrm{cc}
&=
\dfrac{g^2_0}{2}
\sum_{pq}
d^{pq}_\lambda
\tilde{\gamma}^\Lambda_{pq}
\sum_{rs}
d^{rs}_\lambda
\tilde{\gamma}^\Lambda_{rs}
\quad.
\label{eq.l2_crp_cc_energy}
\end{align}
Here, $\tilde{\gamma}^\Lambda_{pq}$ resembles the $\Lambda$-dependent components of the CC response 1-RDM  (\textit{cf.} Eq.\eqref{eq.response_1rdm}), \textit{i.e.}, without $t^a_i$ in $\gamma^\Lambda_{ai}$. In case the cluster and $\Lambda$-operators are truncated to single and double excitations (CRP-CCSD), the first contribution to the CRP-CC correlation energy in Eq.\eqref{eq.l0_crp_cc_energy} reads explicitly
\begin{align}
\Delta\mathcal{V}^{\mathcal{H}}_\mathrm{ccsd}
&=
\sum_{aibj}
\left(
t^a_i
t^b_j
+
t^{ab}_{ij}
\right)
L_{iajb}
\label{eq.l0_crp_ccsd_energy_explicit}
\vspace{0.2cm}
\\
&
+
2
g^2_0
\sum_{aibj}
\left(
t^a_i
t^b_j
+
t^{ab}_{ij}
\right)
d^{ai}_\lambda
d^{bj}_\lambda
\nonumber
\vspace{0.2cm}
\\
&
-
g^2_0
\sum_{aibj}
\left(
t^a_i
t^b_j
+
t^{ab}_{ij}
\right)
d^{bi}_\lambda
d^{aj}_\lambda
\quad,
\nonumber
\end{align}
where we restrict the discussion to closed-shell systems with $L_{iajb}=2g_{iajb}-g_{ibja}$\cite{helgakerbook}. In the first line, we formally have the canonical CCSD energy, however, evaluated with respect to DSE-relaxed amplitudes, while the second and third line contain DSE-induced corrections. The second correlation energy contribution in Eq.\eqref{eq.l1_crp_cc_energy} is given by 
\begin{align}
\Delta\mathcal{V}^d_\mathrm{ccsd}
&=
-
2
g^2_0
\sum_{ai}
d^{ai}_\lambda
t^a_i
\sum_{bj}
d^{bj}_\lambda
t^b_j
\quad,
\label{eq.l1_crp_ccsd_energy_explicit}
\end{align}
and cancels the symmetric disconnected doubles contribution to the DSE correction in the second line of $\Delta\mathcal{V}^{\mathcal{H}}_\mathrm{ccsd}$. The remaining $\Lambda$-dependent correlation contribution in Eq.\eqref{eq.l2_crp_cc_energy}, $\Delta\mathcal{V}^\Lambda_\mathrm{ccsd}$, is evaluated with respect to matrix elements given in Eq.\eqref{eq.response_1rdm}. We close this section by noting, that CRP constraints in Eq.\eqref{eq.crp_constraints} manifest in case of both mean-field and CC theory consistently in cancellation of factorizing (disconnected) DSE contributions.

\subsection{Implementation}
\label{sec.implement}
A pilot implementation of CRP-HF and CRP-CCSD methods (besides its linearised approximations discussed in Sec.\ref{sec.linear_schemes}) was realized via the Python-based Simulations of Chemistry Framework (PySCF) package\cite{sun2018,sun2020}. As shown in Sec.\ref{subsec.crp_cc_equaitons}, the nonlinear CRP-CC working equations can be formulated in analogy to their canonical CC counterparts by absorbing the $\Lambda$-dependence into the one-electron part of the Hamiltonian.
\RestyleAlgo{ruled}
\begin{algorithm}[hbt!]
\caption{CRP-CC Self-Consistency Cycle}
\label{alg.crp_cc_scf}
\textbf{Initialize:} Energy threshold $\Delta\mathcal{V}_\mathrm{min}$/max. iter. $i_\mathrm{max}$
\vspace{0.1cm}
\\
Solve mean-field linearized equations
\vspace{0.1cm}
\\
$\quad\braket{\Phi_\nu\vert e^{-\hat{T}}\{\hat{\mathcal{H}}^\mathrm{crp}_e\}e^{\hat{T}}\vert\Phi_0}=0$ for $\{t_\nu\}$
\vspace{0.1cm}
\\
$\quad\braket{\Phi_0\vert(1+\hat{\Lambda})[e^{-\hat{T}}\{\hat{\mathcal{H}}^\mathrm{crp}_e\}e^{\hat{T}},\hat{\tau}_\nu]\vert\Phi_0}=0$ for $\{\lambda_\nu\}$
\vspace{0.2cm}
\\
Calculate
\vspace{0.1cm}
\\
$\quad d^\Lambda_\lambda=\sum_{rs}d^{rs}_\lambda\gamma^\Lambda_{rs}$ via Eq.\eqref{eq.response_1rdm}
\vspace{0.1cm}
\\
$\quad\Delta\mathcal{V}^{(ec)}_\mathrm{cc}$ via Eq.\eqref{eq.exact_crp_cc_energy}
\vspace{0.2cm}
\\
\For{$0<i\leq i_\mathrm{max}$}{
\vspace{0.2cm}
2a) $\{\mathcal{\hat{H}}^{\Lambda_i}_e\}$ via Eq.\eqref{eq.lambda_eff_fock}
\vspace{0.1cm}
\\
2b) Solve Eqs.\eqref{eq.amplitudes_crp_cc} and \eqref{eq.multipliers_crp_cc} for $\{t^i_\nu,\lambda^i_\nu\}$
\vspace{0.1cm}
\\
2c) $\Delta\mathcal{V}^{(ec)}_\mathrm{cc}(i)$ via Eq.\eqref{eq.exact_crp_cc_energy}
\vspace{0.1cm}
\\
\eIf{$\vert\Delta\mathcal{V}^{(ec)}_\mathrm{cc}(i)-\Delta\mathcal{V}^{(ec)}_\mathrm{cc}(i-1)\vert<\Delta\mathcal{V}_\mathrm{min}$}
    {\vspace{0.2cm}
        CRP-CC correlation energy $\Delta\mathcal{V}^{(ec)}_\mathrm{cc}$ converged
    }
    {\vspace{0.2cm}
    $d^{\Lambda_i}_\lambda=\sum_{rs}d^{rs}_\lambda\gamma^{\Lambda_i}_{rs}$ via Eq.\eqref{eq.response_1rdm} $\Rightarrow$ 2a) $\{\mathcal{\hat{H}}^{\Lambda_{i+1}}_e\}$
    }
}
\end{algorithm}

In scheme \ref{alg.crp_cc_scf}, we illustrate a self-consistent cycle of the coupled CRP-CC problem, which has to be solved iteratively. Thus, the CRP-CC approach is numerically more expensive than canonical CC methods but effectively encodes the optimization of the CBO electronic ground state energy in cavity coordinate space on a CC level of theory. We eventually note that all results discussed in this study were obtained with standard Gaussian atomic orbital basis sets in line with previous work.\cite{fischer2024}

\section{$\Lambda$-Linearisation Schemes}
\label{sec.linear_schemes}
We will now turn to approximations of the CRP-CC approach, which linearize the CRP-CC Lagrangian with respect to $\Lambda$-multipliers. Resulting approximate schemes are formally similar to canonical CC theory and mitigate higher numerical cost of the nonlinear CRP-CC approach.
\begin{table*}[hbt!]
\setlength\extrarowheight{5pt}
\begin{tabular}{l c c l c c c c c}
\hline\hline
$\Lambda$-Linearised CRP-CC Lagrangian &&& \quad Amplitude Equations &&& $t_\nu$ && $\Delta\mathcal{V}^{(ec)}_\mathrm{cc}$\vspace{0.1cm}\\
\hline
\vspace{0.1cm}
$
\mathcal{L}^{\mathrm{mf}}_\mathrm{cc}
=
\mathcal{V}^{(ec)}_\mathrm{rhf}
+
\braket{
\Phi_0
\vert
(
1
+
\hat{\Lambda}
)
e^{-\hat{T}}
\{\hat{\mathcal{H}}^\mathrm{crp}_e\}
e^{\hat{T}}
\vert
\Phi_0}
$
&&&
$
\tilde{R}^{\mathrm{mf}}_\nu
=
\braket{
\Phi_\nu
\vert
e^{-\hat{T}}
\{\hat{\mathcal{H}}^\mathrm{crp}_e\}
e^{\hat{T}}
\vert
\Phi_0}
=
0
$ 
&&&
\textcolor{red}{\xmark}
&&
\textcolor{red}{\xmark}
\vspace{0.1cm}\\         
$
\mathcal{L}^{\Lambda_0}_\mathrm{cc}
=
\mathcal{L}^{\mathrm{mf}}_\mathrm{cc}
-
\dfrac{g^2_0}{2}
\braket{
\Phi_0
\vert
\{\hat{d}^{(e)}_\lambda\}_T
\vert
\Phi_0}^2
$
&&&
$
\tilde{R}^{\Lambda_0}_\nu
=
\tilde{R}^{\mathrm{mf}}_\nu
=
0
$ 
&&&
\textcolor{red}{\xmark}
&&
\textcolor{blue}{\cmark}
\vspace{0.1cm}\\          
$
\mathcal{L}^{\Lambda}_\mathrm{cc}
=
\mathcal{L}^{\Lambda_0}_\mathrm{cc}
-
g^2_0
\braket{
\Phi_0
\vert
\hat{\Lambda}
\{\hat{d}^{(e)}_\lambda\}_T
\vert
\Phi_0}
\braket{
\Phi_0
\vert
\{\hat{d}^{(e)}_\lambda\}_T
\vert
\Phi_0}
$
&&&
$
\tilde{R}^{\Lambda}_\nu
=
\tilde{R}^{\Lambda_0}_\nu
-
g^2_0
\braket{
\Phi_\nu
\vert
\{\hat{d}^{(e)}_\lambda\}_T
\vert
\Phi_0}
\braket{
\Phi_0
\vert
\{\hat{d}^{(e)}_\lambda\}_T
\vert
\Phi_0}
=
0
$ 
&&&
\textcolor{blue}{\cmark}
&&
\textcolor{blue}{\cmark}
\vspace{0.1cm}\\ 
\hline\hline
\end{tabular}
\caption{$\Lambda$-Linearised CRP-CC (lCRP-CC) Lagrangians and corresponding amplitude equations obtained by approximation of $\mathcal{L}^{(ec)}_{\Lambda^2}$ in Eq.\eqref{eq.lambda_squared_lagrangian} with similarity-transformed normal-ordered electronic dipole operator $\{\hat{d}^{(e)}_\lambda\}_T$ defined in Eq.\eqref{eq.short_simtrans_normord_dipole}. Approximations neglect (\textcolor{red}{\xmark}
) or account for (\textcolor{blue}{\cmark}) correlation corrections of the stationary cavity coordinate with respect to the approximate CRP-CC ground state, $t_\nu$, and correlation energy, $\Delta\mathcal{V}^{(ec)}_\mathrm{cc}$.}
\label{tab.lin_lagrange_amplitude}
\end{table*}
We start the discussion by inspecting the last term of the normal-ordered CRP-CC Lagrangian in Eq.\eqref{eq.normal_ordered_crp_cc_lagrangian}, which can be expanded as\cite{Lambda2term}
\begin{align}
\mathcal{L}^{(ec)}_{\Lambda^2}
&=
\dfrac{g^2_0}{2}
\braket{
\Phi_0
\vert
\{\hat{d}^{(e)}_\lambda\}_T
\vert
\Phi_0}^2
\label{eq.lambda_squared_lagrangian}
\vspace{0.2cm}
\\
&+
g^2_0
\braket{
\Phi_0
\vert
\hat{\Lambda}
\{\hat{d}^{(e)}_\lambda\}_T
\vert
\Phi_0}
\braket{
\Phi_0
\vert
\{\hat{d}^{(e)}_\lambda\}_T
\vert
\Phi_0}
\nonumber
\vspace{0.2cm}
\\
&
+
\dfrac{g^2_0}{2}
\braket{
\Phi_0
\vert
\hat{\Lambda}
\{\hat{d}^{(e)}_\lambda\}_T
\vert
\Phi_0}^2
\quad,
\nonumber
\end{align}
where we introduced a short-hand notation for the similarity transformed normal-ordered electronic dipole operator
\begin{align}
\{\hat{d}^{(e)}_\lambda\}_T
&=
e^{-\hat{T}}
\{\hat{d}^{(e)}_\lambda\}
e^{\hat{T}}
\quad.
\label{eq.short_simtrans_normord_dipole}
\end{align}
The last term of Eq.\eqref{eq.lambda_squared_lagrangian} is quadratic in $\Lambda$-multipliers and thus imposes the already discussed coupling between CRP-CC equations. A hierarchy of approximations for $\mathcal{L}^{(ec)}_\mathrm{cc}$ is obtained by linearizing $\mathcal{L}^{(ec)}_{\Lambda^2}$ with respect to $\Lambda$-multipliers following similar ideas from implicit solvation CC theory\cite{cammi2009,caricato2012}. Resulting $\Lambda$-linearized CRP-CC (lCRP-CC) schemes are subject to approximately decoupled CRP-CC working equations, which mimic their canonical CC counterparts and are therefore numerically cheaper to solved. Conceptually, linearisation of the CRP-CC Lagrangian with respect to $\Lambda$-multipliers approximates energy optimization in cavity coordinate space by neglecting correlation corrections of the stationary cavity coordinate.

In Tab.\ref{tab.lin_lagrange_amplitude}, we present $\Lambda$-linearised Lagrangians and related amplitude equations along information on correlation corrections as accounted for in the ground state ($t_\nu$) or energy ($\Delta\mathcal{V}^{(ec)}_\mathrm{cc}$) for three approximation schemes. 

The simplest linearisation scheme entirely neglects $\mathcal{L}^{(ec)}_{\Lambda^2}$, which approximates the stationary cavity coordinate on a mean-field level of theory with Lagrangian, $\mathcal{L}^\mathrm{mf}_\mathrm{cc}$. We  accordingly denote this approach as mean-field linearisation with amplitude equations, $\tilde{R}^{\mathrm{mf}}_\nu$ (\textit{cf.} Tab.\ref{tab.lin_lagrange_amplitude}), as determined by the normal-ordered Hamiltonian $\{\mathcal{\hat{H}}^\mathrm{crp}_e\}$ defined in Eq.\eqref{eq.crp_eff_hamilton}. The only difference between mean-field linearised amplitude equations and their canonical CC counterparts lies in the integral expressions accounting for the electronic DSE contribution. The corresponding mean-field lCRP-CCSD correlation energy reads
\begin{align}
\Delta\mathcal{V}^{\mathrm{mf}}_\mathrm{ccsd}
=
\Delta\mathcal{V}^{\mathcal{H}}_\mathrm{ccsd}
\quad,
\label{eq.mean_field_lin_corr_energy}
\end{align}
which is simply the first term of the full CRP-CCSD correlation energy in Eq.\eqref{eq.exact_crp_cc_energy}. Here, we neglect correlation effects related to the stationary cavity coordinate in both the CC wave function (in terms of amplitudes) and correlation energy as indicated by the two right-most columns of Tab.\ref{tab.lin_lagrange_amplitude}. 

A correction of the correlation energy is obtained by retaining the first term of Eq.\eqref{eq.lambda_squared_lagrangian}, which gives
\begin{align}
\Delta\mathcal{V}^{\Lambda_0}_\mathrm{ccsd}
=
\Delta\mathcal{V}^{\mathcal{H}}_\mathrm{ccsd}
+
\Delta\mathcal{V}^d_\mathrm{ccsd}
\quad,
\label{eq.l0_lin_corr_energy}
\end{align}
for the related $\Lambda_0$-lCRP-CC Lagrangian, $\mathcal{L}^{\Lambda_0}_\mathrm{cc}$. Here, we only correct the energy but rely on the mean-field linearised amplitude equations.

Finally, in order to account for corrections of both correlation energy and amplitudes, we neglect only the nonlinear third term of $\mathcal{L}^{(ec)}_{\Lambda^2}$ in Eq.\eqref{eq.lambda_squared_lagrangian}, which leads to the $\Lambda$-lCRP-CC Lagrangian, $\mathcal{L}^\Lambda_\mathrm{cc}$, as given in the last line of Tab.\ref{tab.lin_lagrange_amplitude}. The correlation energy is here again given by
\begin{align}
\Delta\mathcal{V}^\Lambda_\mathrm{ccsd}
&=
\Delta\mathcal{V}^{\Lambda_0}_\mathrm{ccsd}
\quad,
\label{eq.lambda_lin_corr_energy}
\end{align}
however, the amplitude equations acquire now an additional term (\textit{cf.} Tab.\ref{tab.lin_lagrange_amplitude}), which results from the second term in Eq.\eqref{eq.lambda_squared_lagrangian} linear in $\Lambda$.

\section{Results and Discussion}
\label{sec.num_exp}
We will now discuss CRP-HF, lCRP-CCSD and CRP-CCSD methods for selected molecular model systems\cite{pavosevic2022} namely a Menshutkin reaction between pyridine and methyl bromide (CH$_3$Br) as well as selected methanol-water clusters. In Sec.\ref{subsec.cavity_reorient}, we first address cavity-induced molecular reorientation effects based on a recent proposal by Schnappinger and Kowalewski\cite{schnappinger2024struc}, which provides a molecule-specific unique choice of cavity polarization axis. In Secs.\ref{subsec.result_menshutkin} and \ref{subsec.result_microsolv}, we discuss differences between mean-field and correlated descriptions as well as linearised and fully self-consistent CRP-CCSD approaches for cavity-modified activation and microsolvation energies in molecular model systems, respectively. All energies reported below have been obtained via a Python-based pilot implementation of CRP methods exploiting the PySCF package\cite{sun2018,sun2020}.
\begin{figure*}[hbt]
\begin{center}
\includegraphics[scale=1.0]{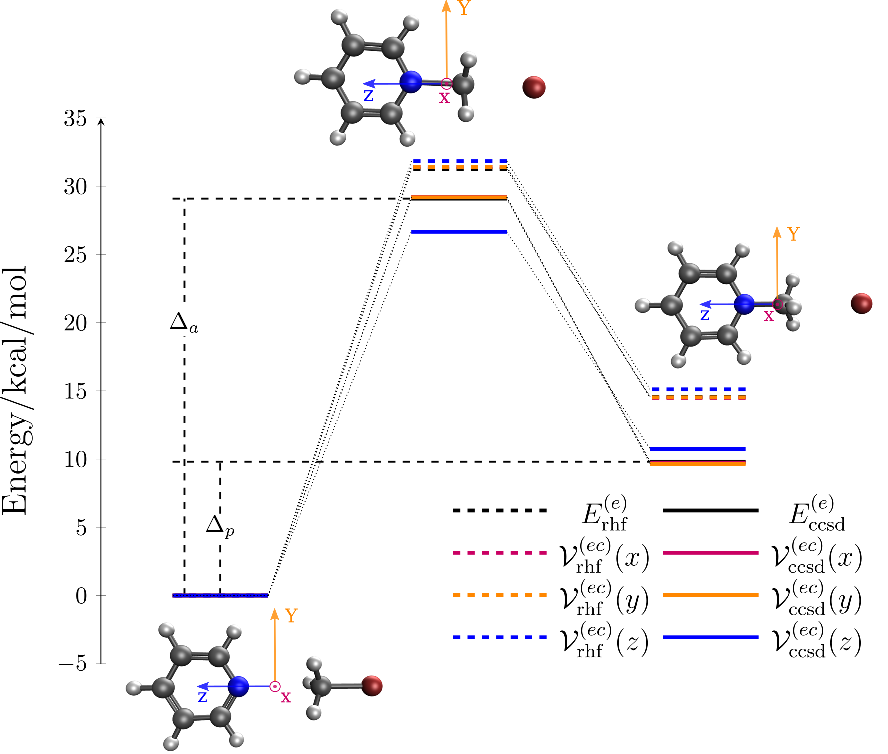}
\end{center}
\renewcommand{\baselinestretch}{1.}
\caption{Cavity-modified electronic energies of the Menshutkin reaction pyridine+CH$_3$Br (hyrodgen in white, carbon in grey, nitrogen in blue, bromine in dark red) at CRP-HF/aug-cc-pVDZ ($\mathcal{V}^{(ec)}_\mathrm{rhf}(\lambda)$, dashed lines) and CRP-CCSD/aug-cc-pVDZ ($\mathcal{V}^{(ec)}_\mathrm{ccsd}(\lambda)$, bold lines) levels of theory with cavity-reoriented reactant (left), transition state (top) and product (right) structures lying in the $y$,$z$-plane ($x$-axis points to the reader). We consider a single cavity mode with polarization $\lambda=x$ (red), $\lambda=y$ (orange) or $\lambda=z$ (blue) and light-matter coupling strength, $g_0=0.03\sqrt{E_h}/ea_0$. The bare electronic reference energies are given by $E^{(e)}_\mathrm{rhf}$ and $E^{(e)}_\mathrm{ccsd}$, whereas activation energy and product energy (illustratively for CCSD energies) are indicated by $\Delta_a$ and $\Delta_p$ (\textit{cf.} Tab.\ref{tab.activation_menshutkin}).}
\label{fig.menshutkin_cavity}
\end{figure*}

Before we proceed, we like to note that our results are not straightforwardly comparable to Ref.\cite{pavosevic2022} from which we extract our model systems since this study relied on \textit{ab initio} QED methods applied in the ESC regime conceptually distinct from correlated \textit{ab initio} CBO methods. For example, correlated \textit{ab initio} QED methods exhibit a cavity frequency-dependency of the ground state energy, which is absent in the CBO framework even in presence of correlation effects.\cite{fischer2024,haugland2025}

\subsection{Cavity-Induced Molecular Reorientation}
\label{subsec.cavity_reorient}
We account for cavity-induced molecular reorientation by transforming the molecular axis system to the principal frame of the molecular polarizability tensor (\textit{cf.} Appendix \ref{subsec.comput_details}). This approach is motivated by an observation reported in Ref.\cite{schnappinger2024struc}, which states that a molecule (in gas phase) under VSC will reorient such that the cavity mode polarization axis (here single mode limit) is aligned with the smallest component of the diagonal polarizability tensor. Accordingly, related eigenvalues and principal axis qualitatively characterize the light-matter interaction strength in connection with corresponding cavity mode polarization axis. For example, the light-matter interaction is minimized along the $x$-axis, which corresponds to the smallest eigenvalue of the polarizability tensor in anisotropic systems with eigenvalues, $\alpha_x<\alpha_y<\alpha_z$, as considered in the following. Additional cavity-induced structural relaxation effects\cite{schnappinger2024struc,liebenthal2024,lexander2024} will not be considered.
We note that in the context of experiments dominantly conducted in condensed phase settings, molecular rotation is hindered such that the cavity mode polarization will never be perfectly aligned to a respective molecular axis. Accordingly, strong coupling effects can be thought of as an average effect taking into account different orientations. In the following, we address cavity-induced modifications from a model perspective for all individual principal axis capturing different limiting orientation scenarios. 

For the Menshutkin reaction between pyridine and CH$_3$Br, we obtain reoriented reactant, transition state and product species as depicted in Fig.\ref{fig.menshutkin_cavity}: The pyridine molecule lies in the $y$,$z$-plane with the reactive N-C-Br-axis being aligned with the $z$-axis, while the $x$-axis points towards the reader. Light-matter interaction is here minimized for a cavity mode polarized perpendicular to the molecular plane ($\lambda=x$) and maximized for a parallel polarization ($\lambda=z$), respectively. We address here cavity-induced modifications of activation and product energies on both mean-field and correlated levels of theory.

\begin{figure*}[hbt]
\begin{center}
\includegraphics[scale=1.0]{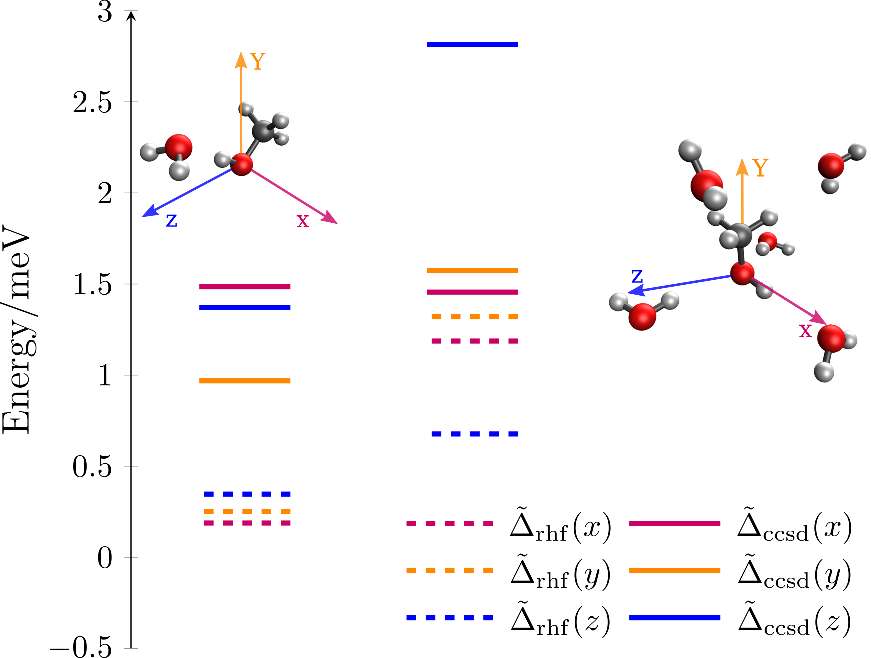}
\end{center}
\renewcommand{\baselinestretch}{1.}
\caption{Cavity-induced modifications of microsolvation energies for cavity-reoriented MeOH@$n$H$_2$O clusters with $n=1$ (left) and $n=5$ (right) (hyrodgen in white, carbon in grey, oxygen in red) at CRP-HF/aug-cc-pVDZ ($\tilde{\Delta}_\mathrm{rhf}(\lambda)$, dashed lines) and CRP-CCSD/aug-cc-pVDZ ($\tilde{\Delta}_\mathrm{ccsd}(\lambda)$, bold lines) levels of theory. We consider a single cavity mode with polarization $\lambda=x$ (red), $\lambda=y$ (orange) or $\lambda=z$ (blue) and light-matter coupling strength, $g_0=0.03\sqrt{E_h}/ea_0$.}
\label{fig.microsolvation_cavity}
\end{figure*}

In case of methanol-water clusters MeOH@$n$H$_2$O with $n=1,5$, we take into account reorientation of the whole cluster treated as one molecular entity (\textit{cf.} Fig.\ref{fig.microsolvation_cavity}). We will address differences in cavity-induced modifications of microsolvation energies on both mean-field and correlated levels of theory and furthermore address collective effects induced by changing the number of water molecules. Notably, we do not account for all potential conformers and related averages in our discussion of strong coupling effects on microsolvation energy, which is beyond the scope of this work.   

\subsection{Cavity-modified Reaction Barriers}
\label{subsec.result_menshutkin}
The Menshutkin reaction between pyridine and CH$_3$Br in gas phase is characterized by reactant, transition state and product species as shown in Fig.\ref{fig.menshutkin_cavity}. In the gas phase scenario, the reaction exhibits a product electronically less stable than reactants, which is significantly stabilized in the presence of solvent effects.\cite{turan2022} 

In a first step, we compare CRP-HF/aug-cc-pVDZ and self-consistent CRP-CCSD/aug-cc-pVDZ results for light-matter coupling strength, $g_0=0.03\sqrt{E_h}/ea_0$, in line with earlier work\cite{fischer2024ir,fischer2024}. We chose an energy convergence threshold of $\Delta\mathcal{V}_\mathrm{min}=10^{-7}\,E_h$ for self-consistent CRP-CCSD calculations in agreement with the underlying CCSD parameter. In Tab.\ref{tab.activation_menshutkin}, we show activation and product energies, $\Delta_a(\lambda)$ and $\Delta_p(\lambda)$, for different cavity mode polarizations, $\lambda$. 
\begin{table}[hbt!]
\caption{Activation energy, $\Delta_a(\lambda)$, and product energy, $\Delta_p(\lambda)$, in kcal/mol as function of cavity mode-polarization, $\lambda$, obtained at CRP-HF/aug-cc-pVDZ and CRP-CCSD/aug-cc-pVDZ levels of theory. Electronic reference energies at a certain level of theory (RHF/CCSD) are equivalent for $\lambda=x,y,z$.}
\begin{center}
\setlength\extrarowheight{3pt}
\begin{tabular}{ll cccccc}
\hline\hline
Method && $\Delta_a(x)$ & $\Delta_a(y)$ &  $\Delta_a(z)$ & $\Delta_p(x)$ & $\Delta_p(y)$ &  $\Delta_p(z)$\\
\hline
RHF       &&   31.24 &  -- & -- & 14.55 & -- & -- \\
CRP-RHF   &&   31.39 &  31.42 & 31.83 & 14.49 & 14.52 & 15.11  \\
CCSD      &&   29.09 &  -- & -- & 9.81 & -- & --  \\
CRP-CCSD  &&   29.20 &  29.18 & 26.66 & 9.73 & 9.65 & 10.75 \\
\hline\hline
\end{tabular}
\end{center}
\label{tab.activation_menshutkin}
\end{table}

At the mean-field level (CRP-RHF), we find relatively small changes due to a presence of the cavity field with magnitudes increasing with cavity polarization direction in agreement with the trend of polarizability tensor eigenvalues. In presence of electron correlation (CRP-CCSD), cavity-induced modifications are especially pronounced along the $z$-axis with a barrier reduction of approximately $2.4\,\text{kcal/mol}$. For all three molecular species, we observe the CRP-CCSD algorithm to converge within one to four iterations, where more iterations are required to converge the energy in case of a $z$-polarized cavity mode responsible for stronger light-matter interaction effects.

Moreover, we find all three lCRP-CCSD schemes to exhibit only minor deviations from the self-consistent CRP-CCSD energy (\textit{cf.} Tab.\ref{tab.comparison_activation_menshutkin}, Appendix \ref{subsec.energy_details}), which provide excellent estimates of cavity-induced electronic energy modifications for the Menshutkin reaction in the single-molecule limit. Due to the hierarchical nature of CRP-CCSD and its linearised variants, this methodology provides a systematic route to systematically addressing the relevance of correlation effects on the stationary cavity coordinate for different molecular systems.

\subsection{Cavity-modified Microsolvation Energies}
\label{subsec.result_microsolv}
As a second example, we consider cavity-induced modifications of microsolvation energies for methanol-water clusters, MeOH@$n$H$_2$O, with $n=1,5$ (\textit{cf.} Fig.\ref{fig.microsolvation_cavity}). Specifically, we address cavity-induced collective effects on both mean-field (CRP-HF) and correlated (CRP-CCSD) levels of theory by comparing two scenarios with different numbers of solvent (water) molecules. 

The cavity-modified electronic energy of a cluster AB can be written as
\begin{align}
\tilde{E}_{AB}
=
\tilde{E}_A
+
\tilde{E}_B
+
\Delta\tilde{E}_{AB}
\quad,
\end{align}
with A=MeOH and B=$n$H$_2$O, respectively. The first two terms correspond to energies of the individual sub-clusters
\begin{align}
\tilde{E}_X
&=
E_X
+
\tilde{\Delta}_X
\quad,
\quad
X
=
A,B
\quad,
\end{align}
with cavity-induced modification, $\tilde{\Delta}_X$. The interaction energy between A and B can be equivalently decomposed as
\begin{align}
\Delta\tilde{E}_{AB}
&=
\Delta E_{AB}
+
\tilde{\Delta}_{AB}
\quad,
\end{align}
where we concentrate in the following on the cavity-induced component, $\tilde{\Delta}_{AB}$, which is roughly two-orders of magnitude smaller than, $\Delta E_{AB}$.

In Fig.\ref{fig.microsolvation_cavity}, we present cavity-induced modifications of microsolvation energies, $\tilde{\Delta}_n(\lambda)$, for $n=1$ (left) and $n=5$ (right) water molecules as function of cavity mode-polarization, $\lambda$. Those result were obtained at CRP-HF/aug-cc-pVDZ and self-consistent CRP-CCSD/aug-cc-pVDZ levels of theory with light-matter coupling strength, $g_0=0.03\,\sqrt{E_h}/ea_0$. In Tab.\ref{tab.cavity_microsolvation}, we provide the corresponding numerical values of $\tilde{\Delta}_n(\lambda)$ given in meV instead of kcal/mol as cavity-induced energy differences are significantly smaller than in the Menshutkin reaction. Note, in the non-interacting limit ($g_0\to0$) all values shown in Tab.\ref{tab.cavity_microsolvation} vanish identically.
\begin{table}[hbt!]
\caption{Cavity-induced modifications of microsolvation energies, $\tilde{\Delta}_n(\lambda)$, in meV for MeOH@$n$H$_2$O with $n=1,5$ as function of cavity mode-polarization, $\lambda$, obtained at CRP-HF/aug-cc-pVDZ and CRP-CCSD/aug-cc-pVDZ levels of theory.}
\begin{center}
\setlength\extrarowheight{3pt}
\begin{tabular}{ll cccccc}
\hline\hline
Method && $\tilde{\Delta}_1(x)$ & $\tilde{\Delta}_5(x)$ &  $\tilde{\Delta}_1(y)$ & $\tilde{\Delta}_5(y)$ & $\tilde{\Delta}_1(z)$ &  $\tilde{\Delta}_5(z)$\\
\hline
CRP-RHF   &&   0.19 & 1.19 &  0.25 & 1.32 & 0.35 & 0.68  \\
CRP-CCSD  &&   1.49 & 1.46 &  0.97 & 1.58 & 1.37 & 2.82 \\
\hline\hline
\end{tabular}
\end{center}
\label{tab.cavity_microsolvation}
\end{table}

In general, we find correlated results to be larger in magnitude than their mean-field equivalents, which illustrates that electron correlation effects can also in this context lead to non-negligible energy modifications under VSC. As before, we chose here an energy convergence threshold of $\Delta\mathcal{V}_\mathrm{min}=10^{-7}\,E_h$ and observe a similar convergence trend, \textit{i.e.},  one to four iterations for $x$- to $z$-polarization axis. Turning to collective effects, we observe significantly stronger differences at the mean-field level (CRP-HF) for $x$- and $y$-polarized cavity modes, whereas for the $z$-polarized scenario a qualitative agreement between CRP-HF and CRP-CCSD results is obtained. We note, a similar finding was reported in Ref.\cite{pavosevic2022} despite the conceptually different methodologies. A numerical comparison of lCRP-CCSD schemes and the self-consistent CRP-CCSD approach reveals also in the present scenario an excellent agreement of energies (\textit{cf.} Tab.\ref{tab.comparison_cavity_microsolvation}, Appendix \ref{subsec.energy_details}). Our results suggest that mean-field results obtained in the CBO framework should be discussed carefully in the context of collective effects in chemical model systems and correlation effects should be taken into account. A future challenge will be related to the question of how correlation effects transition from small local model systems to models of the presumably macroscopic non-local experimental scenario of vibro-polaritonic chemistry.\cite{sidler2020,sidler2025}

\section{Conclusion}
\label{sec.conclusion}
We presented the derivation and implementation of cavity Born-Oppenheimer coupled cluster (CBO-CC) theory in the cavity reaction potential (CRP) framework\cite{fischer2024}, which provides an \textit{ab initio} approach to electron correlation in the vibrational strong coupling regime. The CRP reformulation of the CBO-CC approach satisfies by construction the non-radiating ground state condition and consequently addresses the CBO-CC electronic ground state energy minimized in cavity coordinate space without the need of optimization routines. 

Based on a straightforward CRP-reformulation of CBO-HF theory, we derived the CRP-CC method via a Lagrangian approach formally similar to implicit solvation CC models at the singles and doubles excitation level (CRP-CCSD). The CRP-CCSD Lagrangian is nonlinear in $\Lambda$-multipliers, which results in coupled amplitude and multiplier equations, which have to be solved self-consistently mimicking energy optimization in cavity coordinate space on a CCSD level of theory. Naturally, the CRP-CCSD approach is numerically more expensive than canonical CC theory due to its self-consistent nature. We addressed this bottleneck by introducing a hierarchy of $\Lambda$-linearised CRP-CCSD Lagrangians, which decouple amplitude and multiplier equations by approximating the energy optimization procedure. Resulting lCRP-CCSD methods mimic canonical CC theory and therefore benefit from similar cost. 

We illustratively applied CRP-HF, lCRP-CCSD and CRP-CCSD approaches to two molecular model scenarios: A Menshutkin reaction between pyridine and CH$_3$Br in gas phase besides MeOH@$n$H$_2$O clusters with $n=1,5$. In the first case, we addressed electron correlation effects under VSC on activation and product energies and in the second case we discussed collective effects in cavity-modified microsolvation energies. We find linearised CRP-CCSD methods to provide excellent results compared to the self-consistent CRP-CCSD approach in the few-molecule limit for both scenarios. Furthermore, we find significant cavity-induced electron correlation effects for both model systems and observe collective electronic effects under VSC to differ substantially between correlated and mean-field descriptions. 

The ground state CRP-CCSD approach can be extended to excited states by further exploiting formal similarities with implicit solvation CC theory\cite{cammi2010,caricato2012}. Moreover, the present work provides a natural starting point for the inclusion of implicit solvation effects into \textit{ab initio} vibro-polaritonic chemistry, which would reduce the conceptual gap between theoretical models and experiments dominantly relying on condensed phase settings further. Eventually, the connection between a presumably collective \textit{non-local} nature of vibro-polaritonic chemistry and the inherent relevance of electron correlation on \textit{local} chemical reactions poses another pending open question, whose microscopic side might benefit in the future from the herein presented method development.


\section*{Acknowledgements}
E.W. Fischer acknowledges funding by the Deutsche Forschungsgemeinschaft (DFG, German Research Foundation) through DFG project 536826332 and support by Michael Roemelt. E.W. Fischer thanks Denis Usvyat, Charlotte Rickert, Evelin Christlmaier and Johannes T\"olle for helpful discussions on coupled cluster theory.

\section*{Data Availability Statement}
The data that support the findings of this study are available from the corresponding author upon reasonable request.

\section*{Conflict of Interest}
The author has no conflicts to disclose.

\renewcommand{\thesection}{}
\section*{Appendix}

\setcounter{equation}{0}
\renewcommand{\theequation}{\thesubsection.\arabic{equation}}
\subsection{Derivation of the exact CRP}
\label{subsec.derivation_crp}
We derive the exact CRP in Eq.\eqref{eq.exact_crp}. To this end, we write the CBO electronic Hamiltonian in Eq.\eqref{eq.ground_state_cpes} explicitly for a single cavity mode with polarization, $\lambda$, which leads to
\begin{align}
E^{(ec)}_0
=
\braket{
\Psi^{(ec)}_0
\vert
\hat{H}_e
+
\dfrac{\omega^2_c}{2}
\left(
x_\lambda
-
\dfrac{g_0}{\omega_c}
\hat{d}^{(en)}_\lambda
\right)^2
\vert
\Psi^{(ec)}_0}
\,,
\label{eq.derivation_crp_1}
\end{align}
with polarization-projected molecular dipole operator, $\hat{d}^{(en)}_\lambda=\hat{d}^{(e)}_\lambda+\hat{d}^{(n)}_\lambda$. The stationary cavity coordinate in Eq.\eqref{eq.mincav_exact} will only change the second term of Eq.\eqref{eq.derivation_crp_1}. In a first step, we obtain
\begin{align}
x^0_\lambda
-
\dfrac{g_0}{\omega_c}
\hat{d}^{(en)}_\lambda
&=
\dfrac{g_0}{\omega_c}
\braket{
\Psi^{(ec)}_0
\vert
\hat{d}^{(en)}_\lambda
\vert
\Psi^{(ec)}_0}
-
\dfrac{g_0}{\omega_c}
\hat{d}^{(en)}_\lambda
\,,
\label{eq.derivation_crp_2}
\vspace{0.2cm}
\\
&=
\dfrac{g_0}{\omega_c}
\braket{
\Psi^{(ec)}_0
\vert
\hat{d}^{(e)}_\lambda
\vert
\Psi^{(ec)}_0}
-
\dfrac{g_0}{\omega_c}
\hat{d}^{(e)}_\lambda
\quad,
\label{eq.derivation_crp_3}
\end{align}
where the nuclear dipole operator cancels in Eq.\eqref{eq.derivation_crp_3} due to normalization of the adiabatic ground state. Thus, we find with Eq.\eqref{eq.derivation_crp_1} 
\begin{align}
\mathcal{V}^{(ec)}_0
&=
\braket{
\Psi^{(ec)}_0
\vert
\mathcal{H}_e
\vert
\Psi^{(ec)}_0}
\label{eq.derivation_crp_4}
\vspace{0.2cm}
\\
&-
g^2_0
\braket{
\Psi^{(ec)}_0
\vert
\left(
\braket{
\Psi^{(ec)}_0
\vert
\hat{d}^{(e)}_\lambda
\vert
\Psi^{(ec)}_0}
\hat{d}^{(e)}_\lambda
\right)
\vert
\Psi^{(ec)}_0}
\label{eq.derivation_crp_5}
\vspace{0.2cm}
\\
&+
\dfrac{g^2_0}{2}
\braket{
\Psi^{(ec)}_0
\vert
\hat{d}^{(e)}_\lambda
\vert
\Psi^{(ec)}_0}^2
\quad,
\label{eq.derivation_crp_6}
\end{align}
and obtain Eq.\eqref{eq.exact_crp} by simplifying Eqs.\eqref{eq.derivation_crp_5} and \eqref{eq.derivation_crp_6}. In the first line, we introduced the effective Hamiltonian 
\begin{align}
\mathcal{H}_e
&=
\hat{H}_e
+
\dfrac{g^2_0}{2}
\hat{O}^{(e)}_\lambda
+
\dfrac{g^2_0}{2}
\hat{U}^{(ee)}_\lambda
\quad,
\label{eq.derivation_crp_7}
\end{align} 
whose second quantization representation was given in Eq.\eqref{eq.eff_hamilton}, with one- and two-particle electronic DSE contributions
\begin{align}
\hat{O}^{(e)}_\lambda
&=
e^2
\sum_i
r^2_{i\lambda}
\quad,
\label{eq.origin_invariance_3}
\vspace{0.2cm}
\\
\hat{U}^{(e)}_\lambda
&=
e^2
\sum_{i\neq j}
r_{i\lambda}
r_{j\lambda}
\quad.
\label{eq.origin_invariance_4}
\end{align}

\setcounter{equation}{0}
\renewcommand{\theequation}{\thesubsection.\arabic{equation}}
\subsection{Origin Invariance of the exact CRP}
\label{subsec.origin_invariance_crp}
We show that the CRP, $\mathcal{V}^{(ec)}_0$, is invariant with respect to a shift, $\Delta_\lambda$, of the electronic dipole operator. The latter transforms as
\begin{align}
\hat{d}^{(e)}_\lambda
+
\Delta_\lambda
=
-
e
\sum_i
\left(
r_{i\lambda}
+
\Delta_\lambda
\right)
=
\hat{d}^{(e)}_\lambda
+
N_e
\Delta^{(e)}_\lambda
\,,
\label{eq.origin_invariance_0}
\end{align}
with $\Delta^{(e)}_\lambda=-e\Delta_\lambda$. We simplify the discussion by rewriting Eq.\eqref{eq.exact_crp} as
\begin{align}
\mathcal{V}^{(ec)}_0
&=
\braket{
\Psi^{(ec)}_0
\vert
\hat{H}_e
\vert
\Psi^{(ec)}_0
}
+
\Delta\mathcal{V}^{(ec)}_0
\quad,
\label{eq.origin_invariance_1}
\end{align}
which allows us to consider only the relevant second term
\begin{align}
\Delta\mathcal{V}^{(ec)}_0
&=
\braket{
\hat{O}^{(e)}_\lambda
}
+
\braket{
\hat{U}^{(ee)}_\lambda
}
-
\braket{\hat{d}^{(e)}_\lambda}^2
\quad,
\label{eq.origin_invariance_2}
\end{align} 
with $\braket{\dots}=\braket{\Psi^{(ec)}_0\vert\dots\vert\Psi^{(ec)}_0}$. The three contributions transform under an origin-shift $\Delta_\lambda$ as
\begin{align}
\braket{
\hat{O}^{(e)}_\lambda
}
&\to
\braket{
\hat{O}^{(e)}_\lambda
}
+
2
\braket{
\hat{d}^{(e)}_\lambda
}
\Delta^{(e)}_\lambda
+
N_e
(\Delta^{(e)}_\lambda)^2
\,,
\label{eq.origin_invariance_6}
\vspace{0.2cm}
\\
\braket{
\hat{U}^{(ee)}_\lambda
}
&\to
\braket{
\hat{U}^{(ee)}_\lambda
}
+
2
(
N_e
-
1)
\braket{
\hat{d}^{(e)}_\lambda
}
\Delta^{(e)}_\lambda
\label{eq.origin_invariance_7}
\vspace{0.2cm}
\\
&\hspace{3cm}
+
N_e(N_e-1)
(\Delta^{(e)}_\lambda)^2
\quad,
\nonumber
\vspace{0.2cm}
\\
\braket{\hat{d}^{(e)}_\lambda}^2
&\to
\braket{\hat{d}^{(e)}_\lambda}^2
+
2
N_e
\braket{\hat{d}^{(e)}_\lambda}
\Delta^{(e)}_\lambda
+
N^2_e
(\Delta^{(e)}_\lambda)^2
\,,
\label{eq.origin_invariance_8}
\end{align}
and we note that contributions scaling as $\Delta^{(e)}_\lambda$ and $(\Delta^{(e)}_\lambda)^2$ cancel exactly. Consequently, $\Delta\mathcal{V}^{(ec)}_0$ and $\mathcal{V}^{(ec)}_0$ are origin invariant.

\setcounter{equation}{0}
\renewcommand{\theequation}{\thesubsection.\arabic{equation}}
\subsection{Details on CRP Hartree-Fock Theory}
\label{subsec.derivation_crprhf}
We follow the canonical derivation of HF theory in Ref.\cite{helgakerbook} and expand Eq.\eqref{eq.crp_rhf_energy} in a Taylor series around $\underline{\kappa}=\underline{0}$ as
\begin{align}
\mathcal{V}^{(ec)}_\mathrm{rhf}(\underline{\kappa})
=
\mathcal{V}^{(ec)}_\mathrm{rhf}(\underline{0})
+
\underline{\kappa}^\mathrm{T}
\underline{\mathcal{V}}^{(1)}_\mathrm{rhf}
+
\mathcal{O}(\underline{\kappa}^2)
\quad,
\label{eq.taylor}
\end{align}
where the second term contains the CRP orbital gradient, $\underline{\mathcal{V}}^{(1)}_\mathrm{rhf}$, with elements
\begin{align}
\mathcal{V}^{(1)}_{pq}
=
\left.
\dfrac{\partial\mathcal{V}^{(ec)}_\mathrm{rhf}(\underline{\kappa})}{\partial\kappa_{pq}}
\right\vert_{\underline{\kappa}=\underline{0}}
\quad.
\end{align}
The individual terms of Eq.\eqref{eq.taylor} are conveniently obtained by evaluating the Baker-Campbell-Hausdorff (BCH) expansion of the unitary orbital rotation transformation in Eq.\eqref{eq.crp_rhf_energy} and sort terms in powers of orbital rotation parameters. The first term of Eq.\eqref{eq.crp_rhf_energy} leads to
\begin{align}
\braket{
\Phi^{(ec)}_0
\vert
e^{\hat{\kappa}}
\hat{\mathcal{H}}_e
e^{-\hat{\kappa}}
\vert
\Phi^{(ec)}_0}
&=
\braket{
\Phi^{(ec)}_0
\vert
\hat{\mathcal{H}}_e
\vert
\Phi^{(ec)}_0}
\vspace{0.2cm}
\\
&
+
\braket{
\Phi^{(ec)}_0
\vert
[
\hat{\kappa},
\hat{\mathcal{H}}_e
]
\vert
\Phi^{(ec)}_0}
\nonumber
\vspace{0.2cm}
\\
&
+
\mathcal{O}(\hat{\kappa}^2)
\quad,
\nonumber
\end{align} 
whereas the second term gives
\begin{multline}
\braket{
\Phi^{(ec)}_0
\vert
e^{\hat{\kappa}}
\hat{d}^{(e)}_\lambda
e^{-\hat{\kappa}}
\vert
\Phi^{(ec)}_0}^2
=
\braket{
\Phi^{(ec)}_0
\vert
\hat{d}^{(e)}_\lambda
\vert
\Phi^{(ec)}_0}^2
\\
+
2
\braket{
\Phi^{(ec)}_0
\vert
\hat{d}^{(e)}_\lambda
\vert
\Phi^{(ec)}_0}
\braket{
\Phi^{(ec)}_0
\vert
[
\hat{\kappa},
\hat{d}^{(e)}_\lambda
]
\vert
\Phi^{(ec)}_0}
\\
+
\mathcal{O}(\hat{\kappa}^2)
\quad.
\end{multline}
By collecting terms in powers of $\hat{\kappa}$, we find at zeroth-order 
\begin{align}
\mathcal{V}^{(ec)}_\mathrm{rhf}
&=
\braket{
\Phi^{(ec)}_0
\vert
\hat{\mathcal{H}}_e
\vert
\Phi^{(ec)}_0}
-
\dfrac{g^2_0}{2}
\braket{
\Phi^{(ec)}_0
\vert
\hat{d}^{(e)}_\lambda
\vert
\Phi^{(ec)}_0}^2
\,,
\vspace{0.2cm}
\\
&=
E^{(ec)}_\mathrm{rhf}
+
\dfrac{g^2_0}{2}
\sum_{pq}
O^{pq}_\lambda
D_{pq}
\vspace{0.2cm}
\\
&
\hspace{2cm}
-
\dfrac{g^2_0}{4}
\sum_{pqrs}
d^{ps}_\lambda
d^{rq}_\lambda
D_{pq}
D_{rs}
\nonumber
\quad.
\end{align}
The linear contribution is obtained as
\begin{align}
\mathcal{V}^{(1)}_\mathrm{rhf}
&=
\braket{
\Phi^{(ec)}_0
\vert
[
\hat{\kappa},
\hat{\mathcal{H}}_e
]
\vert
\Phi^{(ec)}_0}
\vspace{0.2cm}
\\
&
\hspace{1cm}
-
g^2_0
\braket{
\Phi^{(ec)}_0
\vert
\hat{d}^{(e)}_\lambda
\vert
\Phi^{(ec)}_0}
\braket{
\Phi^{(ec)}_0
\vert
[
\hat{\kappa},
\hat{d}^{(e)}_\lambda
]
\vert
\Phi^{(ec)}_0}
\nonumber
\,,
\vspace{0.2cm}
\\
&
=
\braket{
\Phi^{(ec)}_0
\vert
[
\hat{\kappa},
\hat{\mathcal{H}}_e
-
g^2_0
\braket{
\hat{d}^{(e)}_\lambda
}_0
\hat{d}^{(e)}_\lambda
]
\vert
\Phi^{(ec)}_0}
\quad,
\vspace{0.2cm}
\\
&
=
\braket{
\Phi^{(ec)}_0
\vert
[
\hat{\kappa},
\hat{\mathcal{H}}^{\Phi_0}_e
]
\vert
\Phi^{(ec)}_0}
\quad,
\end{align}
with
\begin{align}
\braket{
\hat{d}^{(e)}_\lambda
}_0
&=
\braket{
\Phi^{(ec)}_0
\vert
\hat{d}^{(e)}_\lambda
\vert
\Phi^{(ec)}_0}
\quad,
\vspace{0.2cm}
\\
\hat{\mathcal{H}}^{\Phi_0}_e
&=
\hat{\mathcal{H}}_e
-
g^2_0
\braket{
\hat{d}^{(e)}_\lambda
}_0
\hat{d}^{(e)}_\lambda
\quad,
\end{align}
and
\begin{align}
\mathcal{V}^{(1)}_\mathrm{rhf}
=
\sum_{p>q}
\kappa_{pq}
\mathcal{V}^{(1)}_{pq}
\quad,
\end{align}
with $\mathcal{V}^{(1)}_{pq}$ being given by Eq.\eqref{eq.crp_orbital_gradient}, respectively. The CRP Fock operator is subsequently derived from the CRP orbital gradient in the same way as the standard Fock operator.\cite{helgakerbook}.

\setcounter{equation}{0}
\renewcommand{\theequation}{\thesubsection.\arabic{equation}}
\subsection{Derivation of the normal-ordered CRP-CC Lagrangian}
\label{subsec.derivation_normal_crp_cc_lagrange}
We start with $\mathcal{L}^{(ec)}_\mathrm{cc}$ in Eq.\eqref{eq.crp_cbo_cc_lagrangian} and make use of definitions
\begin{align}
\mathcal{\hat{H}}_e
&=
\braket{\mathcal{\hat{H}}_e}_0
+
\{\mathcal{\hat{H}}_e\}
\quad,
\vspace{0.2cm}
\\
\hat{d}^{(e)}_\lambda
&=
\braket{\hat{d}^{(e)}_\lambda}_0
+
\{\hat{d}^{(e)}_\lambda\}
\quad,
\end{align}
with $\braket{\dots}_0=\braket{\Phi^{(ec)}_0\vert\dots\vert\Phi^{(ec)}_0}$, which lead after expansion and rearrangement of terms to
\begin{align}
\mathcal{L}^{(ec)}_\mathrm{cc}
&=
\braket{\hat{\mathcal{H}}_e}_0
-
\dfrac{g^2_0}{2}
\braket{\hat{d}^{(e)}_\lambda}^2_0
\vspace{0.2cm}
\\
&
+
\braket{
\Psi^{(ec)}_\Lambda
\vert
\{\hat{\mathcal{H}}_e\}
\vert
\Psi^{(ec)}_\mathrm{CC}}
\vspace{0.2cm}
\\
&
-
g^2_0
\braket{
\Psi^{(ec)}_\Lambda
\vert
\{\hat{d}^{(e)}_\lambda\}
\vert
\Psi^{(ec)}_\mathrm{CC}}
\braket{\hat{d}^{(e)}_\lambda}_0
\vspace{0.2cm}
\\
&
-
\dfrac{g^2_0}{2}
\braket{
\Psi^{(ec)}_\Lambda
\vert
\{\hat{d}^{(e)}_\lambda\}
\vert
\Psi^{(ec)}_\mathrm{CC}}^2
\quad.
\end{align}
The first line contains the mean-field CRP
\begin{align}
\mathcal{V}^{(ec)}_\mathrm{rhf}
&=
\braket{\hat{\mathcal{H}}_e}_0
-
\dfrac{g^2_0}{2}
\braket{\hat{d}^{(e)}_\lambda}^2_0
\quad,
\end{align}
and from the second and third line, we obtain the effective normal-ordered Hamiltonian
\begin{align}
\hat{\mathcal{H}}^\mathrm{crp}_e
&=
\{\hat{\mathcal{H}}_e\}
-
g^2_0
\braket{\hat{d}^{(e)}_\lambda}_0
\{\hat{d}^{(e)}_\lambda\}
\quad.
\end{align}
The first term expands to
\begin{multline}
\{\mathcal{\hat{H}}_e\}
=
\sum_{pq}
h^{pq}_\lambda
\{\hat{E}_{pq}\}
+
\sum_{pqi}
\bar{w}^{pi}_{qi}
\{\hat{E}_{pq}\}
\\
+
\dfrac{1}{4}
\sum_{pqrs}
\bar{w}^{pq}_{rs}
\{\hat{e}_{pqrs}\}
\quad,
\end{multline}
with
\begin{align}
\bar{w}^{pi}_{qi}
&=
(pi\vert qi)
-
(pi\vert iq)
+
g^2_0
\left(
d^{pq}_\lambda
d^{ii}_\lambda
-
d^{pi}_\lambda
d^{iq}_\lambda
\right)
\quad.
\end{align}
The second term gives
\begin{align}
-
\dfrac{g^2_0}{2}
\braket{\hat{d}^{(e)}_\lambda}_0
\{\hat{d}^{(e)}_\lambda\}
&=
-
g^2_0
\sum_{pqi}
d^{pq}_\lambda
d^{ii}_\lambda
\{\hat{E}_{pq}\}
\quad,
\end{align}
which cancels the symmetric DSE contribution in $\{\mathcal{\hat{H}}_e\}$, such that $\hat{\mathcal{H}}^\mathrm{crp}_e$ in Eq.\eqref{eq.crp_eff_hamilton} is obtained. Collecting the remaining terms finally leads to the normal-ordered CRP-CC Lagrangian in Eq.\eqref{eq.normal_ordered_crp_cc_lagrangian}.

\setcounter{equation}{0}
\renewcommand{\theequation}{\thesubsection.\arabic{equation}}
\subsection{Alternative Representation of the normal-ordered CRP-CC Lagrangian}
\label{subsec.alterantive_normal_crp_cc_lagrange}
We derive Eq.\eqref{eq.lagrangian_crp_energy} from Eq.\eqref{eq.lambda_normal_ordered_crp_cc_lagrangian}. We start from the second term of the latter equation
\begin{align}
\braket{
\Psi^{(ec)}_\Lambda
\vert
\{\hat{\mathcal{H}}^\Lambda_e\}
\vert
\Psi^{(ec)}_\mathrm{CC}}
=
\braket{
\Phi^{(ec)}_0
\vert
e^{-\hat{T}}
\{\hat{\mathcal{H}}^\Lambda_e\}
e^{\hat{T}}
\vert
\Phi^{(ec)}_0}
\vspace{0.2cm}
\\
+
\sum_\nu
\lambda_\nu
\braket{
\Phi^{(ec)}_\nu
\vert
e^{-\hat{T}}
\{\hat{\mathcal{H}}^\Lambda_e\}
e^{\hat{T}}
\vert
\Phi^{(ec)}_0}
\nonumber
\,,
\end{align}
where we already recover the CRP-CC amplitude equations in the second line. The first line turns with the definition of the effective Hamiltonian, $\{\hat{\mathcal{H}}^\Lambda_e\}$, in Eq.\eqref{eq.lambda_crp_eff_hamilton} into
\begin{align}
&
\braket{
\Phi^{(ec)}_0
\vert
e^{-\hat{T}}
\{\hat{\mathcal{H}}^\Lambda_e\}
e^{\hat{T}}
\vert
\Phi^{(ec)}_0}
\vspace{0.2cm}
\\
=&
\braket{
\Phi^{(ec)}_0
\vert
e^{-\hat{T}}
\{\hat{\mathcal{H}}^\mathrm{crp}_e\}
e^{\hat{T}}
\vert
\Phi^{(ec)}_0}
\nonumber
\vspace{0.2cm}
\\
&-
g^2_0
\braket{
\Phi^{(ec)}_0
\vert
e^{-\hat{T}}
\{\hat{d}^{(e)}_\lambda\}
e^{\hat{T}}
\vert
\Phi^{(ec)}_0}^2
\nonumber
\vspace{0.2cm}
\\
&
-
g^2_0
\braket{
\Phi^{(ec)}_0
\vert
\hat{\Lambda}
\{\hat{d}^{(e)}_\lambda\}_T
\vert
\Phi^{(ec)}_0}
\braket{
\Phi^{(ec)}_0
\vert
\{\hat{d}^{(e)}_\lambda\}_T
\vert
\Phi^{(ec)}_0}
\nonumber
\,,
\end{align}
where the third line exploits the short-hand notation introduced in Eq.\eqref{eq.short_simtrans_normord_dipole}. The third term in Eq.\eqref{eq.lambda_normal_ordered_crp_cc_lagrangian} has been explicitly discussed in Sec.\ref{sec.linear_schemes}, and we find that the third term of our latter result is canceled by the second term of Eq.\eqref{eq.lambda_squared_lagrangian}. When we furthermore take into account the first and third term of Eq.\eqref{eq.lambda_squared_lagrangian}, we arrive at
\begin{align}
\mathcal{L}^{(ec)}_\mathrm{cc}
&=
\mathcal{V}^{(ec)}_\mathrm{rhf}
+
\braket{
\Phi^{(ec)}_0
\vert
e^{-\hat{T}}
\{\hat{\mathcal{H}}^\mathrm{crp}_e\}
e^{\hat{T}}
\vert
\Phi^{(ec)}_0}
\vspace{0.2cm}
\\
&-
\dfrac{g^2_0}{2}
\braket{
\Phi^{(ec)}_0
\vert
e^{-\hat{T}}
\{\hat{d}^{(e)}_\lambda\}
e^{\hat{T}}
\vert
\Phi^{(ec)}_0}^2
\vspace{0.2cm}
\\
&
+
\dfrac{g^2_0}{2}
\braket{
\Phi_0
\vert
\hat{\Lambda}
e^{-\hat{T}}
\{\hat{d}^{(e)}_\lambda\}
e^{\hat{T}}
\vert
\Phi_0}^2
\vspace{0.2cm}
\\
&+
\sum_\nu
\lambda_\nu
\braket{
\Phi^{(ec)}_\nu
\vert
e^{-\hat{T}}
\{\hat{\mathcal{H}}^\Lambda_e\}
e^{\hat{T}}
\vert
\Phi^{(ec)}_0}
\quad,
\end{align}
and finally arrive at Eq.\eqref{eq.lagrangian_crp_energy} by defining (\textit{cf.} Eq.\eqref{eq.exact_crp_cc_energy})
\begin{align}
\mathcal{V}^{(ec)}_\mathrm{cc}
&=
\mathcal{V}^{(ec)}_\mathrm{rhf}
+
\braket{
\Phi^{(ec)}_0
\vert
e^{-\hat{T}}
\{\hat{\mathcal{H}}^\mathrm{crp}_e\}
e^{\hat{T}}
\vert
\Phi^{(ec)}_0}
\vspace{0.2cm}
\\
&-
\dfrac{g^2_0}{2}
\braket{
\Phi^{(ec)}_0
\vert
e^{-\hat{T}}
\{\hat{d}^{(e)}_\lambda\}
e^{\hat{T}}
\vert
\Phi^{(ec)}_0}^2
\vspace{0.2cm}
\\
&
+
\dfrac{g^2_0}{2}
\braket{
\Phi_0
\vert
\hat{\Lambda}
e^{-\hat{T}}
\{\hat{d}^{(e)}_\lambda\}
e^{\hat{T}}
\vert
\Phi_0}^2
\quad.
\end{align}

\setcounter{equation}{0}
\renewcommand{\theequation}{\thesubsection.\arabic{equation}}
\subsection{Computational Details}
\label{subsec.comput_details}
All molecular structures were taken from Ref.\cite{pavosevic2022}. In order to reorient the molecules, we first obtained the diagonal polarizability tensor and the corresponding eigenvectors via ORCA 5.0.4\cite{orca2012,orcav52022} on both RHF/aug-cc-pVDZ and DLPNO-CCSD/aug-cc-pVDZ levels of theory. Since molecular orientation becomes essentially a new structural parameter in cavity scenarios, it will depend on the level of theory employed to obtain the diagonal polarizability tensor. We chose the latter here consistent with the respective CRP electronic structure method. The matrix of eigenvectors provides a coordinate transformation to the polarizability tensor's principal axis frame, which transforms initial molecular structures to their cavity-reorientated equivalents. CRP-HF, lCRP-CCSD and CRP-CCSD methods were subsequently applied.

\setcounter{equation}{0}
\renewcommand{\theequation}{\thesubsection.\arabic{equation}}
\setcounter{table}{0}
\renewcommand{\thetable}{\thesubsection.\Roman{table}}
\subsection{Details on Cavity-Modified Electronic Energies}
\label{subsec.energy_details}
In Tab.\ref{tab.comparison_activation_menshutkin}, we compare cavity-induced modifications, $\tilde{\Delta}_x(\lambda)$, of activation and product energies for the Menshutkin reaction 
\begin{align}
\tilde{\Delta}_x(\lambda)
=
\Delta_x(\lambda)
-
E^{(e)}_\mathrm{ccsd}
\quad,
\quad
x
=
a,p
\quad,
\end{align}
with respect to the three linearised CRP-CCSD schemes (\textit{cf.} Tab.\ref{tab.lin_lagrange_amplitude}) and the self-consistent CRP-CCSD approach.
\begin{table}[hbt!]
\caption{Cavity-induced changes of activation energy, $\tilde{\Delta}_a(\lambda)$, and product energy, $\tilde{\Delta}_p(\lambda)$, in kcal/mol as function of cavity mode-polarization, $\lambda$, obtained for different lCRP-CCSD/aug-cc-pVDZ schemes and self-consistent CRP-CCSD/aug-cc-pVDZ. }
\begin{center}
\setlength\extrarowheight{3pt}
\begin{tabular}{ll cccccc}
\hline\hline
Method && $\tilde{\Delta}_a(x)$ & $\tilde{\Delta}_a(y)$ &  $\tilde{\Delta}_a(z)$ & $\tilde{\Delta}_p(x)$ & $\tilde{\Delta}_p(y)$ &  $\tilde{\Delta}_p(z)$\\
\hline
mf-lCRP-CCSD            &&   0.1164 & 0.0906 & -2.4242 & -0.0722 & -0.1606 & 0.9424  \\
$\Lambda_0$-lCRP-CCSD   &&   0.1164 & 0.0906 & -2.4259 & -0.0722 & -0.1606 & 0.9406  \\
$\Lambda$-lCRP-CCSD     &&   0.1160 & 0.0902 & -2.4252 & -0.0725 & -0.1608 & 0.9416  \\
CRP-CCSD                &&   0.1158 & 0.0900 & -2.4301 & -0.0706 & -0.1590 & 0.9391  \\
\hline\hline
\end{tabular}
\end{center}
\label{tab.comparison_activation_menshutkin}
\end{table}

In Tab.\ref{tab.comparison_cavity_microsolvation}, we show a similar comparison for cavity-induced modifications of microsolvation energies, $\tilde{\Delta}_n(\lambda)$, of MeOH@$n$H$_2$O with $n=1,5$
\begin{table}[hbt!]
\caption{Cavity-induced modifications of microsolvation energies, $\tilde{\Delta}_n$, in meV for MeOH@$n$H$_2$O with $n=1,5$ as function of cavity mode-polarization, $\lambda$, obtained for different lCRP-CCSD/aug-cc-pVDZ schemes and self-consistent CRP-CCSD/aug-cc-pVDZ.}
\begin{center}
\setlength\extrarowheight{3pt}
\begin{tabular}{ll cccccc}
\hline\hline
Method && $\tilde{\Delta}_1(x)$ & $\tilde{\Delta}_5(x)$ &  $\tilde{\Delta}_1(y)$ & $\tilde{\Delta}_5(y)$ & $\tilde{\Delta}_1(z)$ &  $\tilde{\Delta}_5(z)$\\
\hline
mf-lCRP-CCSD            &&   1.4844 & 1.4514 &  0.9838 & 1.5863 & 1.3624 & 2.8113 \\
$\Lambda_0$-lCRP-CCSD   &&   1.4842 & 1.4560 &  0.9763 & 1.5837 & 1.3666 & 2.8102 \\
$\Lambda$-lCRP-CCSD     &&   1.4834 & 1.4479 &  0.9778 & 1.5795 & 1.3669 & 2.8151 \\
CRP-CCSD                &&   1.4850 & 1.4556 &  0.9685 & 1.5752 & 1.3708 & 2.8156 \\
\hline\hline
\end{tabular}
\end{center}
\label{tab.comparison_cavity_microsolvation}
\end{table}


%


\begin{thebibliography}{200}
\bibitem{Lambda2term} We note that $\mathcal{L}^{(ec)}_{\Lambda^2}$ enters Eq.\eqref{eq.normal_ordered_crp_cc_lagrangian} with a minus sign.
\end{thebibliography}
\end{document}